\documentclass[aps,prl,twocolumn,floats,letterpaper,superscriptaddress,tighten]{revtex4}

\usepackage{amssymb}
\usepackage{amsbsy}
\usepackage{amsmath}
\usepackage{graphicx}
\usepackage{graphics}
\usepackage{setspace}
\usepackage{array}
\usepackage{color}
\usepackage{xcolor}
\usepackage{fontenc}
\usepackage{textcomp}
\usepackage{rotating}
\usepackage{bm}
\usepackage{hyperref}
\hypersetup{pdfpagemode=UseNone}
\usepackage{subfigure}
\usepackage{chngcntr}

\DeclareMathOperator{\Tr}{Tr}
\DeclareMathOperator{\sgn}{sgn}

\newcommand{\e}{\varepsilon}

\newcommand{\vex}[1]{\bm{\mathrm{#1}}}

\newcommand{\pup}[1]{{\scriptscriptstyle{({#1})}}}

\newcommand{\ket}[1]{| {#1} \rangle}

\newcommand{\tauh}{\hat{\tau}}

\newcommand{\sigh}{\hat{\sigma}}

\newcommand{\bsub}{\begin{subequations}}
\newcommand{\esub}{\end{subequations}}
\newcommand{\ts}[1]{{\textstyle{{#1}}}}

\newcommand{\be}{\begin{equation}}
\newcommand{\ee}{\end{equation}}
\newcommand{\bea}{\begin{eqnarray}}
\newcommand{\eea}{\end{eqnarray}}

\newcommand{\taucol}{\tau_{\scriptscriptstyle{\mathsf{col}}}}

\newcommand{\intl}[1]{\int\limits_{#1}}
\newcommand{\rbz}{\mathrm{RBZ}}

\begin{document}
\title{Fractionalization Waves in Two-dimensional Dirac Fermions: \\
Quantum Imprint from One Dimension}
\author{Seth M.\ Davis}
\affiliation{Department of Physics and Astronomy, Rice University, Houston, Texas 77005, USA}
\author{Matthew S.\ Foster}
\affiliation{Department of Physics and Astronomy, Rice University, Houston, Texas 77005, USA}
\affiliation{Rice Center for Quantum Materials, Rice University, Houston, Texas 77005, USA}
\date{\today}

\begin{abstract}
Particle fractionalization is believed to orchestrate the physics of many strongly correlated systems, 
yet its direct experimental detection remains a challenge. We propose a simple measurement for an 
ultracold matter system, in which correlations in initially decoupled 1D chains are imprinted via quantum quench 
upon two-dimensional Dirac fermions. Luttinger liquid correlations launch relativistic ``fractionalization waves'' 
along the chains, while coupling noninteracting chains induces perpendicular dispersion. 
These could be easily distinguished in an ultracold gas experiment.
\end{abstract}
\maketitle


Fractionalization is a profound, nonperturbative effect of interparticle interactions in quantum matter, 
in which the emergent degrees of freedom of a strongly correlated system can be neither bosonic nor fermionic. 
Fractionalization may reside at the heart of high-temperature superconductivity and spin liquid physics \cite{Lee06,Subir08}. 
Although an essential characteristic of the fractional quantum Hall effect (FQHE) 
that may enable topological quantum computation \cite{Nayak08},
the direct detection of fractionalization in solid state experiments has proven 
to be challenging \cite{Willett09}.  

In this Letter, we propose an ultracold fermion gas experiment that could detect a clear signal for fractionalization,
using currently available experimental techniques. In comparison to the much more daunting task of realizing 
a FQHE state of interacting fermions in cold atoms \cite{ColdAtomFQHE}, we require only the preparation of 1D fermionic Luttinger liquids 
(as recently measured in \cite{ColdAtom-LL1,ColdAtom-LL2}), 
that can be coupled together via a quantum quench \cite{Esslinger15}
into a 2D pi-flux lattice (as recently realized in \cite{ColdAtom-piflux1,ColdAtom-piflux2,ColdAtom-piflux3}).
We predict discriminating signatures in density waves launched from an initial Gaussian bump at the time
of the quench. Any degree of fractionalization produces waves with a characteristic shape profile
that propagate at the ``speed of light'' along the 1D chains. By contrast, a noninteracting prequench system 
induces simple dispersion perpendicular to the chains.

\begin{figure}[b!]
\includegraphics[angle=0,width=.42\textwidth]{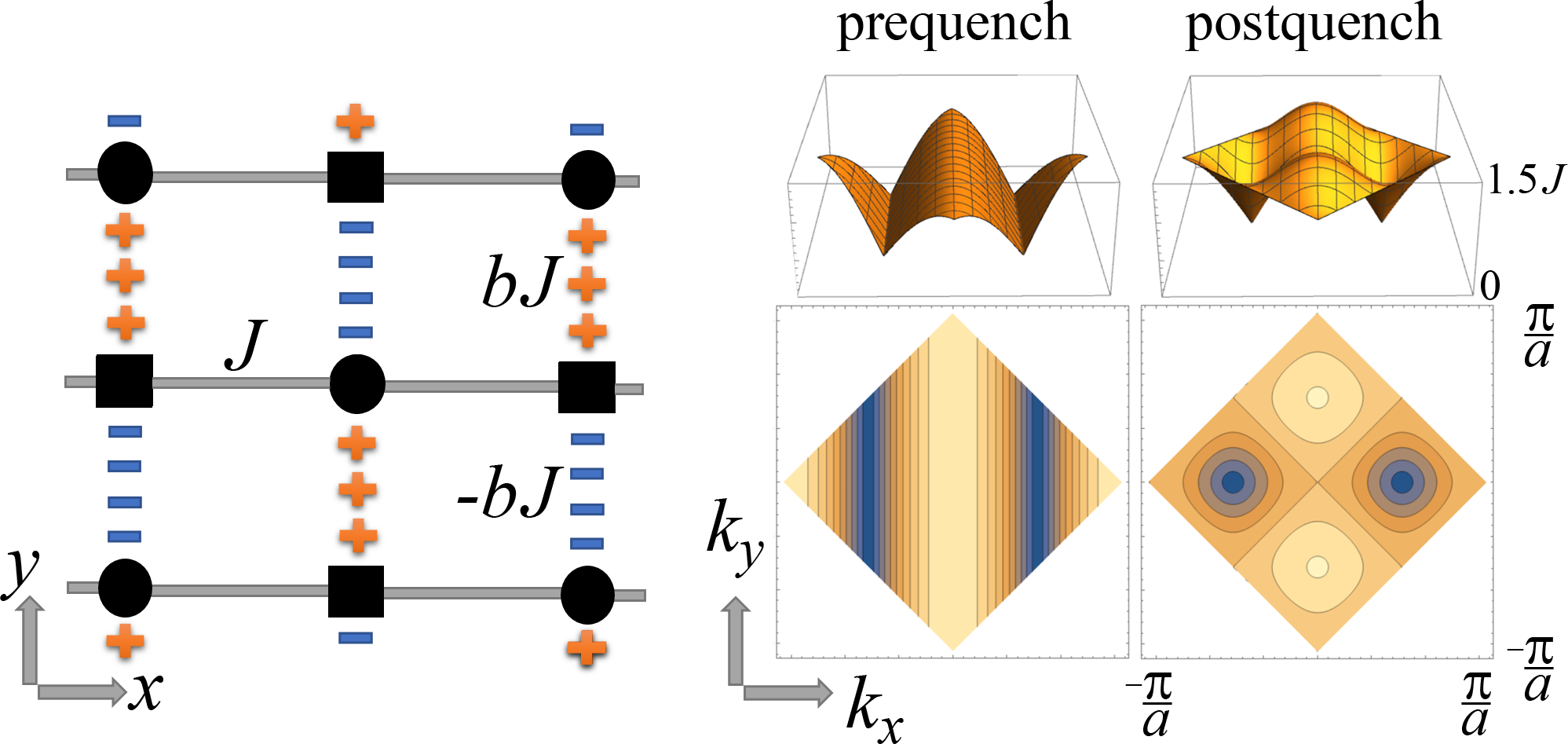}
\caption{Lattice setup for quench-induced fractionalization waves. 
We consider fermions hopping on a pi-flux square lattice (left), with horizontal bonds of strength $J$ 
and sign-staggered vertical bonds of strength $b J$. The Hamiltonian for the noninteracting model is given 
by Eq.~(\ref{LatHam}). 
We quench from the decoupled chain limit ($b = 0$) to $b > 0$. We assume that interactions can induce Luttinger 
liquid correlations (charge fractionalization) along the chains in the prequench state. The latter are imprinted by 
the quench upon the pi-flux band fermions, leading to the two-dimensional density wave dynamics depicted in 
Fig.~\ref{Fig--FracQuench} (fractionalized) and Fig.~\ref{Fig--NoFracQuench} (not fractionalized, initially noninteracting).
The right panels in this figure show the pre- and postquench energy bands. The prequench state of 
decoupled chains is characterized by vertical nodal lines. The postquench band gaps these out, except for a pair 
of Dirac points at $\{k_x,k_y\} = \{\pm \pi/2a,0\}$. 
Energy bands are depicted over the reduced Brillouin zone of the pi-flux lattice; $a$ denotes the lattice constant. 
} 
\label{Fig--QuenchGeom}
\end{figure}

\begin{figure}[b!]
\includegraphics[angle=0,width=.38\textwidth]{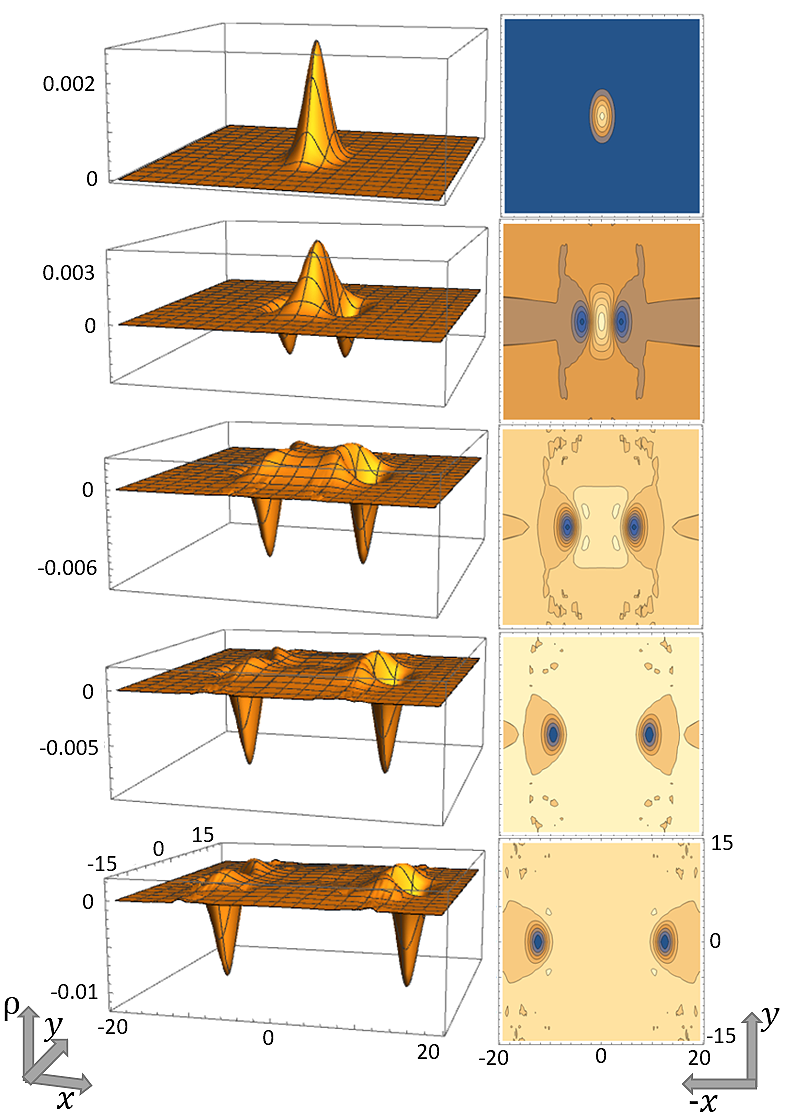}
\caption{Quench from decoupled chains to the pi-flux lattice I:
horizontal, ``relativistic'' fractionalization waves. 
A positive Gaussian density bump is superimposed by an external potential on top of decoupled Luttinger liquids in the initial state. 
Via instantaneous quench, the interchain coupling is turned on, while the bump is released by turning off the potential. 
In this figure we plot the time evolution of the post-quench density profile $\rho(t,x,y)$,
assuming charge fractionalization of the initial Luttinger liquid. 
Left (right) panels show the density in profile (contour) plots; \emph{negative} density
means a depletion of the filled Fermi sea. Total particle number is conserved \cite{SM}. 
The above plots give time-slice profiles at $t = 0,3,6,9,12$ (where $t$ is in units of 
$v_F/a$). 
The degree of fractionalization is characterized by the fermion anomalous dimension $\eta = 0.7$ [Eq.~(\ref{IC-Corr})]. 
We also incorporate a small but positive initial Fermi momentum, $k_F =0.1/a$. 
The other initial configuration parameters are 
\{$Q = 0.1,\Delta_x = 2a,\Delta_y = 3a$\} [Eq.~(\ref{PhiDef})], 
and we time-evolve with the Hamiltonian parameters 
\{$v_F = b = 1$\}. 
We choose a small bump to conserve computational resources, 
but in an experiment a larger bump would minimize lattice-scale detail neglected in Eq.~(\ref{IC-Corr}) \cite{SM}}.
\label{Fig--FracQuench}
\end{figure}

\begin{figure}[t!]
\includegraphics[angle=0,width=.365\textwidth]{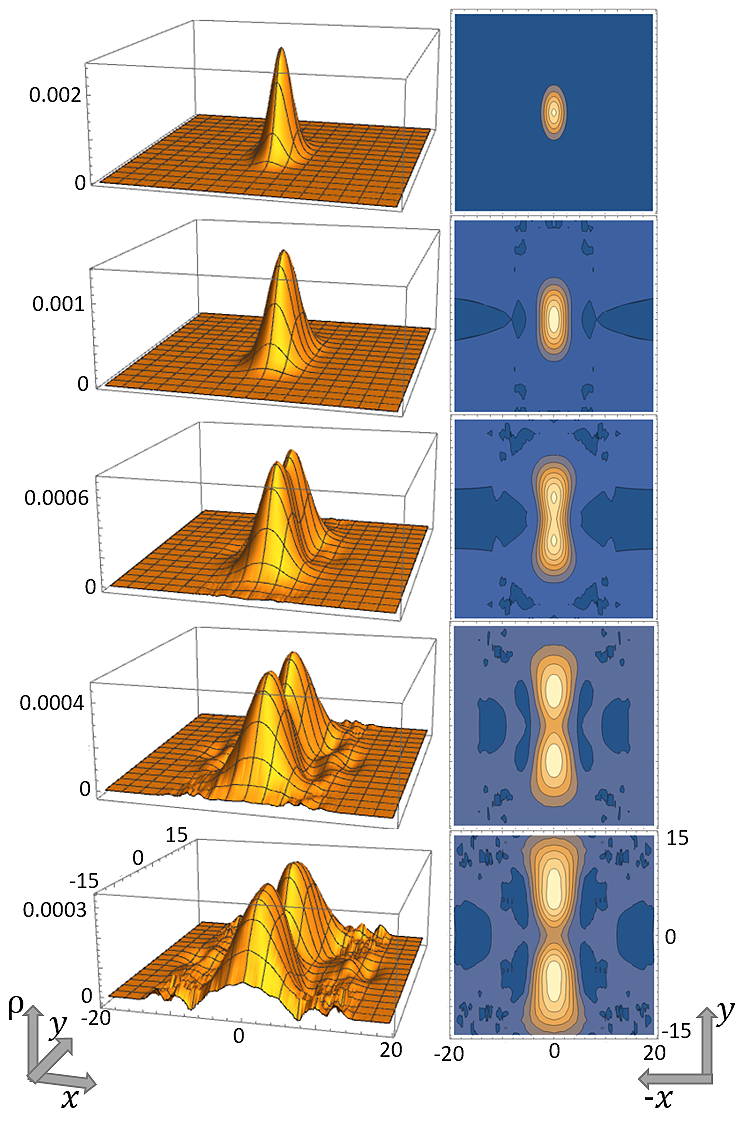}
\caption{Quench from decoupled chains to the pi-flux lattice II:
dispersing vertical density waves from non-fractionalized (noninteracting) chains.  
The absence of ``relativistic'' propagation in the horizontal direction is due to Pauli blocking (see text and Fig.~\ref{Fig--Wig7}). 
We plot the same time evolution of the post-quench density profile as in Fig.~\ref{Fig--FracQuench}, but for a vanishing prequench fermion anomalous dimension $\eta = 0$. All other parameters are identical to 
Fig.~\ref{Fig--FracQuench}.  
The plots show profiles at $t = 0,3,6,9,12$.}
\label{Fig--NoFracQuench}
\end{figure}

Fractionalization can arise in a many-fermion system when the fermion operator acquires an anomalous dimension,
due to interactions \cite{Cardy,Sachdev-Book}. In our proposed experiment,
a nonzero fermion anomalous dimension
directly determines density wave dynamics in a two-dimensional (2D) fermion system.
Here Luttinger liquid correlations 
\cite{Tsvelik,Giamarchi}
in a system of initially decoupled 1D chains
are \emph{imprinted} upon two-dimensional Dirac fermions. This is accomplished via a quantum quench 
\cite{Cazalilla06,Kehrein08,Werner09,Rosch12,Mitra13,Mirlin13,Konik15,White16,Radzihovsky16,MitraRev}
that couples together the chains into a 2D pi-flux lattice model (see Fig.~\ref{Fig--QuenchGeom}).
To probe the dynamics, we calculate the density waves emitted from an initial density bump
\cite{Bettelheim06,Foster10,Foster11,Mossel10,Lancaster1,Lancaster2,Marquardt12,Lancaster3,Fagotti16,Doyon17,Kormos17}. 
We show that a nonzero initial-state-fermion anomalous dimension launches relativistic 
``fractionalization waves'' along the chains, shown in Fig.~\ref{Fig--FracQuench}. 
By contrast, the same quench performed from initially noninteracting chains induces 
dispersive propagation perpendicular to the chains, see Fig.~\ref{Fig--NoFracQuench}. 
The key result of this work is that 
the orthogonal motions of the fractionalized and noninteracting cases 
should be easily distinguishable in an ultracold fermion experiment.

Similar fractionalization waves dubbed ``supersolitons'' were previously predicted in 
1D quenches, including the continuum sine-Gordon model \cite{Foster10} and the \emph{XXZ} chain \cite{Foster11}. 
As in those studies, correlations shape the initial condition, but we ignore interactions in the post-quench evolution.
In an ultracold fermion gas it might be possible to turn off the interactions at the time of the quench, 
but this is not a requirement for us. 
Interactions are strongly irrelevant (in the sense of the renormalization group) in the post-quench band structure. 
Our results should hold over a tunable transient window $0 \leq t < \taucol$, where
$1/\taucol$ is the particle-particle scattering rate determined by the interactions and the post-quench energy density.


\textit{Model}.---We consider a pi-flux lattice model for fermions in 2D, 
\begin{align}\label{LatHam}
	\!\!\!\!
	H 
	=
	-
	J
	\sum_{m,n}
	c^\dagger_{m,n}
	\!
	\left[	
	c_{m+1,n}
	+
	b
	\,
	(-1)^{m+n}
	c_{m,n+1}
	\right]
	+
	\text{H.c.},
	\!\!\!\!
\end{align}
where $c_{m,n}$ annihilates a fermion at site $\{x,y\} = \{m,n\}a$ of the square lattice,
$a$ is the lattice spacing, 
and $J > 0$ is the hopping energy.
The dimensionless anisotropy parameter $b$ controls the strength of the staggered vertical hopping
(see Fig.~\ref{Fig--QuenchGeom}). 
We work with spinless fermions without loss of generality.
Spin-1/2 particles would be advantageous in an ultracold fermion experiment, as 
decoupled Hubbard chains in the prequench state
give a particular way to realize tunable charge fractionalization via the on-site Hubbard $U$ interaction 
(at densities away from half-filling) \cite{Giamarchi}.
Apart from this, physical spin will not impact the dynamics discussed here.

The system is assumed to initially have $b = 0$, so that 
the lattice reduces to a set of uncoupled 1D chains. 
The noninteracting band structure consists of vertical nodal lines. 
Via instantaneous quantum quench, $b$ is switched to a positive, 
nonzero value, gapping out the nodal lines except for a pair of Dirac points at $\{k_x,k_y\} = \{\pm \pi/2a,0\}$.
The pi-flux ensures that the low-energy sectors of the pre- and post-quench band structures overlap.

We assume that the main effect of the global quench is to excite particle-hole 
pairs along the nodal lines of the prequench band structure. We therefore retain
momentum modes along narrow channels including these, 
$\{|k_x \pm \pi/2a| \leq \Lambda,|k_y| \leq \pi/2a\}$. Here $\Lambda \ll \pi/a$ is 
a momentum cutoff. 
Eq.~(\ref{LatHam}) can then be approximated as 
\bsub
\begin{align}
	\label{Heff}
	H 
	\simeq&\,
	v_F 
	\int_{-\Lambda}^\Lambda
	\frac{d k_x}{2 \pi}
	\int_{-\frac{\pi}{2a}}^{\frac{\pi}{2a}}
	\frac{d k_y}{2 \pi}
	\,
	\psi^\dagger(\vex{k})
	\,
	\hat{h}(\vex{k})
	\,
	\psi(\vex{k})
\nonumber\\
	&\,
	+
	\int
	d x \, d y\, 
	\psi^\dagger(x,y) 
	\,
	\psi(x,y)
	\,
	\Phi(x,y)	
	+
	H_I,
\\
	\label{h}
	\hat{h}(\vex{k})
	\equiv&\,
	\sigh^3 
	\,
	k_x
	+
	\sigh^2
	\,
	m(k_y),
\end{align}
\esub
where $v_F = 2 J a$ is the maximum band velocity. 
The field $\psi(\vex{k})\rightarrow \psi_{\sigma,\tau}(\vex{k})$ is a four-component spinor.
The Pauli matrices $\sigh^{1,2,3}$ act on the space of 
right ($\sigma^3 = 1$) and left ($\sigma^3 = -1$) movers in the initial decoupled chains; this is \emph{not} equivalent
to the space of right and left nodal lines (see Ref.~\cite{SM} for details). 
Eq.~(\ref{Heff}) is invariant under SU(2) rotations on the index $\tau \in \{1,2\}$,
which distinguishes modes whose $k_y$ momenta fall in or outside
the \emph{reduced} Brillouin zone (RBZ) depicted in Fig.~\ref{Fig--QuenchGeom}.
The parameter 
\begin{align}\label{mDef}
	m(k_y)
	\equiv
	(b/a) 
	\sin(k_y a) 
\end{align}
in Eq.~(\ref{h})
plays the role of a $k_y$-dependent ``mass,'' when the system is viewed as a collection of decoupled
1D chains. Linearizing near $k_y = 0$ with $b = 1$ would give isotropic massless 2D Dirac fermions.

Relative to the homogeneous, noninteracting lattice model in Eq.~(\ref{LatHam}), we have 
incorporated two additional perturbations on the second line of Eq.~(\ref{Heff}).
The first is an inhomogeneous external potential $\Phi(x,y)$. We assume a localized
potential profile in the prequench state. Via the axial anomaly, 
this induces an initial density inhomogeneity in the decoupled chains of the form 
$\rho(x,y) = - \kappa \, \Phi(x,y)$, where $\kappa$ is the compressibility. 
After the quench we will set $\Phi = 0$, and we will monitor the evolution of 
$\rho(t,x,y)$ as a probe of the dynamics. 

Although Eq.~(\ref{LatHam}) with $b = 0$ describes decoupled chains for spinless fermions, 
the field $\psi_{\sigma,\tau}$ consists of four, not two components. The synthetic $\tau$-spin degree
of freedom is an artifact of folding into the RBZ, necessary for describing the pi-flux lattice.
The term $H_I$ in Eq.~(\ref{Heff}) encodes generic short-ranged intrachain fermion-fermion interactions. 
It is well-known that the low-energy theory for a single-channel, spin-1/2 
SU(2)-symmetric
quantum wire
(described via a four-component field $\psi$)
admits four independent
local, four-fermion interaction operators \cite{Giamarchi}. This includes spin current-current and
charge umklapp interactions, and these can gap out the spin or charge degrees of freedom. 
In our case the coupling constants of these operators should be tuned precisely to zero,
because these describe interactions between \emph{pairs} of chains. 
The admissible interactions [charge current-current and U(1) stress tensor operators] set
the charge velocity and Luttinger parameter $K$ in the prequench Luttinger liquid state of
the decoupled chains \cite{SM}.   

At time $t = 0$, we quench on the interchain coupling $b > 0$, and turn off the potential $\Phi$ and interactions in $H_I$. 
(In fact, we argue later that interactions can remain in place, and will produce a negligible effect on the dynamics up to time $\taucol$, defined below). 
Then the time-evolving density profile is determined by the convolution
\begin{align}\label{rho(t)}
	\rho(t,x,y) 
	=&\,
	\int d x_1 \, d x_2 \, d y_1 \, 
	\Tr\left[
		\hat{G}^\dagger(t,x_1,y_1)
		\,
		\hat{G}(t,x_2,y_1)
	\right]
	\nonumber\\
	&\,
	\times \mathcal{C}_\Phi(x - x_1,x - x_2; y - y_1),
\end{align}
where $\hat{G}(t,x_1,y_1)$ is the causal Green's function associated to $\hat{h}$, and 
$\mathcal{C}_\Phi(x_1,x_2;y)$ describes the static one-particle fermion correlation function 
in the initial state at linear order in the external potential $\Phi$, given by \cite{Foster11,SM}
\begin{align}\label{IC-Corr}
	\!
	\mathcal{C}_{\Phi}(x_1,x_2;y) 
	= 
	\frac{c_\eta}{2}
	\!
	\left[\frac{\alpha^2}{(x_{12})^2 + \zeta^2}\right]^{\eta/2}
	\!
	\frac{
	\int_{x_1}^{x_2} 
	d x \,
	\kappa
	\,
	\Phi(x,y)
	}{x_{12}},
	\!
\end{align}
where $x_{12} = x_1 - x_2$. 
Here $c_\eta$ and $\alpha$ are positive constants, while
$\eta$ is the fermion anomalous dimension. 
The latter is $\eta = (1/2^p)(K + K^{-1} - 2)$, 
where $K$ is the Luttinger parameter \cite{Tsvelik,Giamarchi}.
The exponent $p = 1$ ($p = 2$) for spinless (spin-1/2) fermions within each chain. 
$K = 1$ gives $\eta = 0$ (noninteracting chains); otherwise $\eta > 0$. 
The parameter $\zeta$ is a short-distance regularization that can affect the dynamics at 
long times \cite{Foster11}. Eq.~(\ref{IC-Corr}) is appropriate for half-filling ($k_F = 0$). 
We compare the results from this correlator with an exact lattice quench in the free fermion case in Ref.~\cite{SM}.

\textit{Results}.---We numerically integrate Eq.~(\ref{rho(t)}), using (\ref{IC-Corr}) and assuming a
Gaussian potential
\begin{align}\label{PhiDef}
	\kappa \, \Phi(x,y) = Q \, (\pi \Delta_x \Delta_y)^{-1} \, e^{-\left({x}/{\Delta_x}\right)^2-\left({y}/{\Delta_y}\right)^2}.
\end{align}
We set the parameter $\alpha = a = 1$ in Eq.~(\ref{IC-Corr}) \cite{SM}. 

As is clear from Figs.~\ref{Fig--FracQuench} and \ref{Fig--NoFracQuench},
the qualitative behavior of the post-quench density profile depends crucially on whether or not the initial state is 
fractionalized. In the initially fractionalized case [$\eta > 0$ in Eq.~(\ref{IC-Corr}), Fig.~\ref{Fig--FracQuench}], 
the density develops collective excitations that propagate horizontally along the chains at the 
maximum band velocity $v_F$. 
These fractionalization waves (``supersolitons'' \cite{Foster10,Foster11}) retain their shape as they travel and exhibit 
power-law growth of amplitude with time. Supersolitons possess positive peaks and negative-density troughs;
the total particle number induced by the initial potential on top of the filled Fermi sea is preserved at all times \cite{SM}. 
The results in Fig.~\ref{Fig--FracQuench} obtain from Eq.~(\ref{IC-Corr}) with no short-distance regularization, $\zeta = 0$.
Nonzero $\zeta$ can arise due to the effects of irrelevant operators \cite{Foster11}, 
but we show in Ref.~\cite{SM} that qualitatively identical dynamics obtain in this case except at long times, wherein the supersoliton growth
is curtailed \cite{Foster11}. 
By contrast, for the noninteracting initial condition  ($\eta = 0$, Fig.~\ref{Fig--NoFracQuench}),
there is no supersoliton and the initial density disperses vertically, perpendicular to the chains.

The density dynamics in Figs.~\ref{Fig--FracQuench} and \ref{Fig--NoFracQuench} should be contrasted with one-particle quantum mechanics. 
The same Green's function $\hat{G}(t,x,y)$ that enters into Eq.~(\ref{rho(t)}) determines the evolution of a Gaussian single-particle wavepacket. 
Since $\hat{h} \simeq -i \sigh^3 \partial_x - i \sigh^2 \partial_y$, the result is a circular wavefront propagating at the ``speed of light'' 
\cite{SM}. Instead, the fractionalized quench gives $x$-directed supersolitons, while the noninteracting quench gives $y$-dispersing 
propagation. The difference between the latter and one-particle quantum mechanics is due to Pauli blocking \cite{Foster11}.    
Single-particle quantum mechanics also shows that lattice-scale detail neglected in the Green's function 
has negligible effect on the dynamics over the time scales of interest \cite{SM}.

\begin{figure}[t!]
	\includegraphics[angle=0,width=.28\textwidth]{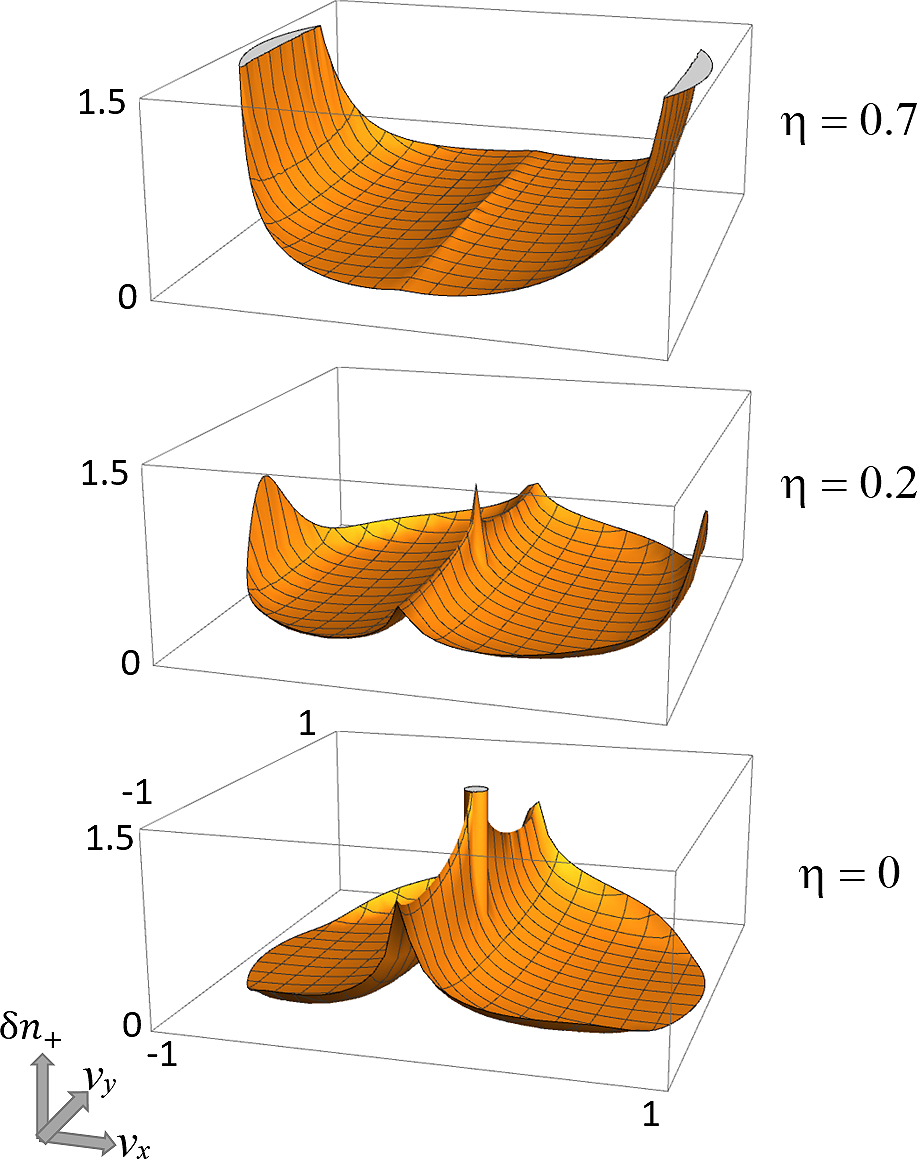}
\caption{Wigner velocity distribution $\delta n_{+}(v_x,v_y;R_x,R_y)$  imprinted on the pi-flux band fermions at the time of the quench, 
induced by charge fractionalization and a Gaussian density bump in the initial state of decoupled chains. 
The distribution is evaluated at the center of the bump $R_x = R_y = 0$.
Unlike continuum 2D massless Dirac fermions, 
the allowed velocities span a disk due to the lattice regularization in the $k_y$-direction, Eqs.~(\ref{Heff}) and (\ref{VDisk}).
The parameter $\eta$ is the fermion anomalous dimension [Eq.~(\ref{IC-Corr})]. 
For $\eta > 0$, there is a divergence of the $x$-velocities near the ``speed of light'' (band velocity $v_F = 1$).
At $\eta = 0$ (noninteracting initial condition), large $x$-velocities are suppressed by Pauli blocking (see text). 
}
\label{Fig--Wig7}
\end{figure}

Further insight into the quench dynamics obtains from the Wigner distribution due to the initial Gaussian bump,
\begin{align}\label{Wigner}
	\delta n_{+}(v_x,v_y;R_x,&\,R_y)
	\propto
	\mathcal{J}(\vex{v}) 
	\int
	d^2 \vex{q}
	\,
	e^{i \vex{q}\cdot\vex{R}}
\nonumber\\
	\times&
	\left\langle
		a^\dagger\left[\vex{k}(\vex{v}) - {\vex{q}}/{2}\right]
		a\left[\vex{k}(\vex{v}) + {\vex{q}}/{2}\right]
	\right\rangle_\Phi,
\end{align}
where $a(\vex{k})$ annihilates a pi-flux conduction band fermion with momentum $\vex{k}$, 
and 
$\mathcal{J}(\vex{v}) \equiv \left|{\partial k_\mu }/{\partial v_\nu}\right|$ is the Jacobian 
relating the post-quench band velocities 
$v_{x,y} \equiv \partial_{k_{x,y}} \sqrt{k_x^2 + m^2(k_y)}$ 
to the momenta $k_{x,y}$. 
The expectation value in Eq.~(\ref{Wigner}) is computed
at time $t = 0$, i.e.\ using Eq.~(\ref{IC-Corr}). 
The Wigner distribution is plotted for variable $\eta$ in Fig.~\ref{Fig--Wig7}. 
We observe a pronounced difference between the fractionalized and non-fractionalized cases. 
The distribution for the case with $\eta > 0$ has a diverging density of $x$-direction 
velocities approaching the maximum band velocity. On the other hand, the noninteracting
$\eta = 0$ case has a velocity distribution strongly localized to $x$-velocities close to zero. 
The absence of large $x$-velocities in the latter is due to Pauli blocking: 
for $k_y = 0$,
the momentum $|q_x|$ must exceed $2 |k_x|$
in order for the pair of operators in Eq.~(\ref{Wigner}) to create a particle-hole pair in the
Fermi sea. Large $|q_x| > 1/\Delta_x$ (and therefore large $v_x$) is suppressed by the initial density profile [Eq.~(\ref{PhiDef})]. 
Fractionalization ($\eta > 0$) circumnavigates the Pauli blocking restriction on large $x$-velocities \cite{SM}.
This is because the fermions responsible for the post-quench propagation are not locally related to the 
effectively noninteracting, but fractionally charged constituents that define the pre-quench vacuum state \cite{Foster11}. 
We also note that the 
kinematic condition
\begin{equation}\label{VDisk}
	v_x^2 + v_y^2/b^2 \leq 1
\end{equation}
implies that the accumulation of the Wigner distribution at $v_x = 1$ requires $v_y = 0$, 
and thus explains why the supersoliton is stable to dispersion in the $y$-direction.

\textit{Window for collisionless dynamics}.---The
correlator in Eq.~(\ref{IC-Corr}) with $\eta > 0$ arises
due to generic short-ranged interactions in the initially 
decoupled chains. The post-quench dynamics captured
by Eq.~(\ref{rho(t)}) ignore interactions in the subsequent
time evolution. 

Beyond the density bump dynamics explored here, the main
bulk effect of the global quench is to generate a finite
density of particle-hole pairs, corresponding to a nonzero
average energy per particle. If the interactions are not 
turned off at the time of the quench, then the system
is expected to eventually thermalize to a temperature
corresponding to the injected energy density \cite{Mitra13}. 

We can estimate the collision rate $1/\taucol$ responsible for thermalization 
in the post-quench evolution.
For the low-energy pi-flux Dirac fermions, a short-ranged lattice interaction 
carries units of energy$\times$length${}^2$ $\sim U_f \, a^2$, where $U_f$ is
the lattice interaction energy.  
The subscript ``$f$'' denotes the interaction strength after the quench,
which can differ from the prequench strength $\equiv U_i$. 
The post-quench Fermi's golden rule collision rate should be of order
$
	1/\taucol \sim \left[(U_f \, a^2) / (b v_F^2)\right]^2 \, \e^3,
$
where $\e$ is the characteristic energy 
per particle. 
The two factors of $b^{-1} v_F^{-2}$ arise from the density of states. 
For $k_F = 0$ (half filling), 
a crude estimate is $\e \sim b \, v_F/a \sim b J$, so that
\begin{align}\label{taucol}
	b  J \, \taucol \sim (J/U_f)^2. 
\end{align}
In the post-quench evolution, time is measured in units of $1/b J$ 
($b = 1$ in Figs.~\ref{Fig--FracQuench} and \ref{Fig--NoFracQuench}). 
Eq.~(\ref{taucol}) implies that the window of time over which 
collisionless dynamics can occur is set by the dimensionless ratio of $U_f/J$.
At times $t \gg \taucol$, 
we expect the ultimate density wave evolution to be governed by 
classical hydrodynamics \cite{Rosch12}.

Small $U_f/J$ will induce a large collisionless window. 
By contrast, the anomalous dimension $\eta$ responsible for
the supersoliton dynamics in Fig.~\ref{Fig--FracQuench} 
is a function of the ratio $U_i/J$. 
Taking the latter to be too small will result in $\eta \ll 1$. 
For spin-1/2 Hubbard chains with repulsive interactions and 
$U_i/J \sim 1$, it is possible to get $K$ close to 1/2 ($\eta = 1/8$) 
for particle densities very close but not equal to half-filling 
\cite{Giamarchi,Schulz90}. 
In Ref.~\cite{SM}, we show for example that $\eta = 0.2$ 
still exhibits the supersoliton over the same time interval as
Fig.~\ref{Fig--FracQuench}.

A balance should be struck between minimizing collisions
over a sufficiently long time window after the quench, 
and maximizing the correlations in the initial state.
At the same time, in an optical lattice setup for ultracold fermions $U_f < U_i$,
since lowering the tunneling barriers in the $y$-direction to couple the 1D chains 
together will necessarily ``unsqueeze'' the atoms in that direction, decreasing
the on-site interaction energy.  
A further reduction of $U_f$ and (enhancement of $\taucol$) 
is possible if at the time of the quench, the confinement is simultaneously reduced in 
the $z$-direction, perpendicular to the plane of the 2D lattice.


We thank Kaden Hazzard and Randy Hulet for helpful discussions. 
We thank Stephen Bradshaw,  
Anthony Sciola, 
Jia-Liang Shen, 
and 
Shah Alam for helpful discussions on high-performance computing.
This work was supported in part by the Data Analysis and Visualization Cyberinfrastructure 
funded by NSF under Grant No.~OCI-0959097 and Rice University.
M.S.F.\ acknowledges
support from the U.S. Army Research Office (Grant W911NF-17-1-0259).
This research 
was also supported 
by NSF CAREER Grant No.~DMR-1552327,
and
by the Welch Foundation Grant No.~C-1809. 
M.\ S.\ F.\
thanks the Aspen Center for Physics, which is supported
by the NSF Grant No.~PHY-1607611, for its hospitality
while part of this work was performed.

\newpage \clearpage

\onecolumngrid

\begin{center}
	{\large
	Fractionalization Waves in Two-dimensional Dirac Fermions: \\
	Quantum Imprint from One Dimension
	\vspace{4pt}
	\\
	SUPPLEMENTAL MATERIAL
	}
\end{center}
\counterwithin{figure}{section}
\counterwithin{equation}{section}
\makeatletter
\setcounter{equation}{0}
\setcounter{figure}{0}
\setcounter{table}{0}
\setcounter{page}{1}
\renewcommand{\theequation}{S\arabic{equation}}
\renewcommand{\thefigure}{S\arabic{figure}}
\renewcommand{\bibnumfmt}[1]{[S#1]}
\renewcommand{\citenumfont}[1]{S#1}

\begingroup
\hypersetup{linkbordercolor=white}
\tableofcontents
\endgroup


\section{I.\ Hamiltonian, sublattice operators, and $\tau$-SU(2) pseudospin}

The lattice Hamiltonian in Eq.~(1) of the main text can be rewritten in terms of operators $c_{A,B}$, 
which annihilate fermions on the A and B sublattices. 
These are respectively indicated by squares and circles 
in Fig.~1 (left panel). The Fourier modes of the sublattice operators carry momenta that span the reduced Brillouin zone ($\rbz$). 
Then Eq.~(1) can be expressed as
\begin{align}\label{S--H2}
	H 
	=
	2 J
	\intl{\rbz}
	\frac{d^2\vex{k}}{(2 \pi)^2}
	\Psi^\dagger(\vex{k})
	\left[
	-
	\cos(k_x a) 
	\,
	\sigh^1 
	+
	a
	\,
	m(k_y)
	\,
	\sigh^2
	\right]
	\Psi(\vex{k}),
\qquad
	\Psi(\vex{k})
	\equiv
	\begin{bmatrix}
	c_A(\vex{k})	\\
	c_B(\vex{k})
	\end{bmatrix}.
\end{align}
Here the Pauli matrices $\sigh^{1,2,3}$ act on the sublattice space, 
with $\sigma^3 = \pm 1$ corresponding to sublattices A and B, respectively.
The ``mass'' $m(k_y)$ was defined by Eq.~(3). 
Linearizing Eq.~(\ref{S--H2}) in the vicinity of the right and left nodal lines
of the $b = 0$ prequench Hamiltonian (Fig.~1) 
gives 
\begin{align}\label{S--H3}
	H 
	\simeq 
	v_F 
	\int_{-\Lambda}^\Lambda
	\frac{d k_x}{2 \pi}
	\int_{-\frac{\pi}{2a}}^{\frac{\pi}{2a}}
	\frac{d k_y}{2 \pi}
	\,
	\psi^\dagger(\vex{k})
	\left[
	k_x
	\,
	\sigh^1 \tauh^3
	+
	m(k_y)
	\,
	\sigh^2
	\right]
	\psi(\vex{k}),
\qquad	
	\psi(\vex{k})
	\equiv
	\begin{bmatrix}
	\Psi\left(\vex{k} + \frac{\pi}{2 a} \hat{x}\right)	\\
	\Psi\left(\vex{k} - \frac{\pi}{2 a} \hat{x}\right)
	\end{bmatrix},
\qquad
	v_F = 2 J a. 
\end{align}
$\psi(\vex{k}) \rightarrow \psi_{\sigma,\tau}(\vex{k})$ is now a four-component spinor. 
The Pauli matrices $\tauh^{1,2,3}$ act on the right/left nodal line space,
with $\tau^3 = \pm 1$ corresponding to the nodal lines $k_x = \pm \pi/2a$. 

The final form of the (unperturbed) Hamiltonian on the first and third lines of Eq.~(2) 
obtains from Eq.~(\ref{S--H3}) after a basis rotation, 
wherein 
\begin{align}
	\label{S--psiR}
\quad
	\psi(\vex{k})
	\Rightarrow
	\frac{1}{\sqrt{2}}
	\left(\hat{1} + i \sigh^2 \tauh^3\right)
	\psi(\vex{k})
	=
	\frac{1}{\sqrt{2}}
	\left[
	\begin{aligned}
		&(c_A + c_B)(\vex{k} + \ts{\frac{\pi}{2 a}})	\\
	-\,	&(c_A - c_B)(\vex{k} + \ts{\frac{\pi}{2 a}})	\\
		&(c_A - c_B)(\vex{k} - \ts{\frac{\pi}{2 a}})	\\
		&(c_A + c_B)(\vex{k} - \ts{\frac{\pi}{2 a}})
	\end{aligned}
	\right]
	=
	\left[
	\begin{aligned}
		& R^\pup{i}(\vex{k})	\\
	-\,	& L^\pup{o}(\vex{k})	\\
		& R^\pup{o}(\vex{k})	\\
		& L^\pup{i}(\vex{k})
	\end{aligned}
	\right].
\end{align}
Here, $R^\pup{i}(\vex{k})$ and $R^\pup{o}(\vex{k})$ denote right-movers in the prequench
system of decoupled chains. The ``$i$'' and ``$o$'' superscripts distinguish modes
whose $k_y$-values in the full Brillouin zone $|k_{x,y}| \leq \pi/a$ respectively 
reside inside or outside of the RBZ shown in Fig.~1. 
The shift of the outside modes by a sublattice reciprocal lattice vector into the reduced zone
effectively sends $c_{A,B} \rightarrow \pm c_{A,B}$. 

In the basis defined by Eqs.~(2) and (\ref{S--psiR}), 
the $\sigh^{1,2,3}$ Pauli matrices act on the right/left-mover space, while
the $\tauh^{1,2,3}$ matrices grade the inner/outer mode space.


\section{II.\ Green's functions and time evolution}

The expression for the density $\rho(t,x,y)$ in Eq.~(4) exploits the Heisenberg evolution of the Dirac operator 
$\psi(t,x,y)$ via the noninteracting pi-flux Hamilonian $\hat{h}(\vex{k})$ in Eq.~(2b). 
This is given by the convolution 
\begin{equation}\label{S--psiEvol}
	\psi(t,x,y) 
	= 
	\int d x_1 d y_1 
	\,
	\hat{G}(t,x - x_1,y - y_1)
	\,
	\psi(x_1,y_1),
\end{equation}
where $\psi(x,y)$ is the Schroedinger operator. 
Setting $v_F = 1$ [Eq.~(2)], the causal Green's function is
\begin{align}\label{S--G2+1}
\begin{aligned}[b]
	\hat{G}(t,x,y) 
	=&\,
	i
	\int_{-\infty}^{\infty}\frac{d\omega}{2\pi}
	\int_{-\infty}^{\infty}\frac{dk_x}{2\pi}
	\int_{-\pi/(2a)}^{\pi/(2a)}\frac{dk_y}{2\pi}
	\,
	e^{-i \omega t + i k_x x + i k_y y}
	\left[\frac{\omega + \sigh^3 \, k_x + \sigh^2 \, m(k_y)}{(\omega + i\eta)^2 - (k_x)^2 - m^2(k_y)}\right]
\\
	=&\,
	\int_{-\pi/(2a)}^{\pi/(2a)}\frac{dk_y}{2\pi}
	\,
	e^{i k_y y}
	\,
	\hat{G}^\pup{1+1}[t,x;m(k_y)],
\end{aligned}
\end{align}
where
\bsub\label{S--G1+1Tot}
\begin{align}\label{S--G1+1}
	\hat{G}^\pup{1+1}(t,x;m)
	\equiv&\, 
	i
	\int_{-\infty}^{\infty}
	\frac{d\omega}{2\pi}
	\int_{-\infty}^{\infty}\frac{dk_x}{2\pi}
	\,
	e^{-i \omega  t + i k_x  x}
	\left[\frac{\omega + \sigh^3 \, k_x + \sigh^2 \, m}{(\omega + i\eta)^2 - k_x^2 - m^{2}}\right]
\nonumber\\
	=&\,
	\theta(t)
	\left\{	
		\begin{bmatrix}
		\delta(t-x)	&	0\\
		0		&	\delta(t+x)
		\end{bmatrix}
		+
		\begin{bmatrix}
		G^\pup{1}(t,x;m) 	& G^\pup{2}(t,x;m)\\
		-G^\pup{2}(t,x;m) 	& G^\pup{1}(t,-x;m)
		\end{bmatrix}
		\theta\left(t^2 - x^2\right)
	\right\},
\\
	G^\pup{1}(t,x;m)
	=&\,
	-
	\frac{m}{2} 
	\left(\frac{t + x}{\sqrt{t^2 - x^2}}\right)
	J_1\left(m \sqrt{t^2 - x^2}\right),
	\quad
	G^\pup{2}(t,x;m)
	=
	-
	\frac{m}{2} 
	J_0\left(m \sqrt{t^2 - x^2}\right),
	\label{BesselComp}
\end{align}
\esub
and $\theta(t)$ denotes the Heaviside step function.
In Eq.~(\ref{BesselComp}) $J_{0,1}(z)$ are Bessel functions of the first kind. 
The causal Green's function $\hat{G}^\pup{1+1}(t,x;m)$ in Eqs.~(\ref{S--G2+1}) and (\ref{S--G1+1})
describes the propagation of 1+1-D massive Dirac fermions \cite{S--Foster11}.

We can restore full lattice detail in the $x$-direction by sending $k_x \rightarrow m(k_x)$
[with $b = 1$ in Eq.~(3)].  
In $t$-$\vex{k}$ space, we then have 
\begin{align}\label{S--GFull}
\begin{aligned}[b]
	\hat{G}_{FL}(t,k_x,k_y) 
	=&\,
	i
	\int_{-\infty}^{\infty}\frac{d\omega}{2\pi}
	\,
	e^{-i \omega t }
	\left[\frac{\omega + \sigh^3 \, m(k_x) + \sigh^2 \, m(k_y)}{(\omega + i\eta)^2 - m^2(k_x) - m^2(k_y)}\right]
\\
    =&\,
	\cos\left[t\sqrt{m^2(k_x)+m^2(k_y)}\right] 
	- 
	i\  
	\frac{\sigh^3 m(k_x) + \sigh^2 m(k_y)}{\sqrt{m^2(k_x)+m^2(k_y)}}
	\sin\left[t\sqrt{m^2(k_x)+m^2(k_y)}\right].
\end{aligned}
\end{align}
We will use this form in Sec.~V below to check that our continuum approximation in the $x$-direction 
does not meaningfully impact the dynamics on the time scales of interest.


\section{III.\ Initial state correlation function}

In the initial state $b = 0$, so that Eq.~(2) describes decoupled 1D chains. 
These are perturbed by the external potential $\Phi(x,y)$ and generic
short-ranged \emph{intrachain} interactions [encoded in $H_I$, Eq.~(2a)]. 
The static one-particle correlation function in the prequench state is given by 
\cite{S--Tsvelik,S--Giamarchi,S--Foster11}
\begin{align}\label{S--IC-Corr}
	\left\langle \psi^\dagger_{\sigma_1,\tau_1}(x_1,y_1) \, \psi_{\sigma_2,\tau_2}(x_2,y_2)\right\rangle
	=&\,
	\delta_{\sigma_1,\sigma_2}
	\,
	\delta_{\tau_1,\tau_2}
	\,
	\delta(y_1 - y_2)
	\,
	(-1)^{1 + \sigma_1}
\nonumber\\
	&\,
	\times
	\frac{i c_\eta \alpha^\eta}{2 \pi}
	\frac{\sgn(x_1 - x_2)}{|x_1 - x_2|^{1 + \eta}}
	\exp\left\{
		(-1)^{1 + \sigma_1}
		\,	
		i \pi \kappa
		\int_{x_1}^{x_2} 
		d z \,
		\left[- \Phi(z,y)\right]
	\right\}.
\end{align}
Here $\sigma_1 = \pm 1$ corresponds to right- and left-moving fermions. 
The initial Hamiltonian and the correlator in Eq.~(\ref{S--IC-Corr}) are both independent of the
``inner/outer'' [Eq.~(\ref{S--psiR})] $\tau$-pseudospin space. 
The positive constant $c_\eta$ is defined as \cite{S--Foster11}
\[
	c_\eta \equiv \sqrt{\pi} \Gamma\left({\textstyle{1 + \frac{\eta}{2}}}\right)/\Gamma\left({\textstyle{\frac{1+\eta}{2}}}\right).
\]
However, the overall normalization of Eq.~(\ref{S--IC-Corr}) is not completely determined for $\eta > 0$ due to the 
prefactor $\alpha^\eta$, where $\alpha$ is an ultraviolet length scale that is not defined within the
continuum bosonization method \cite{S--Giamarchi}
(it could be extracted from numerics \cite{S--Foster11}, or in favorable cases via the Bethe ansatz). 
The anomalous dimension $\eta = (1/2^p)(K + K^{-1} - 2)$, where $K$ is the Luttinger parameter \cite{S--Tsvelik,S--Giamarchi,S--Foster11}.
The exponent $p = 1$ ($p = 2$) for spinless (spin-1/2) fermions within the chains. 
The external potential $\Phi(x,y)$ appears in the ``gauge string'' (phase) as consequence of the 1+1-D axial anomaly; 
$\kappa = (\partial n / \partial \mu) = 2^{p-1} K / (\pi u)$ is the compressibility. 
Here $u$ denotes the interaction-renormalized charge velocity.   

Linearizing Eq.~(\ref{S--IC-Corr}) in $\Phi(x,y)$ and discarding the zeroth order term (which does not enter the density dynamics),
we obtain the correlator quoted in Eq.~(5) of the main text if we set $\zeta = 0$ there.
Nonzero $\zeta$ can arise from irrelevant operators; although negligible for long-wavelength, long-time dynamics near equilibrium,
irrelevant operators can have a strong effect for far-from-equilibrium (e.g., quantum quench) time evolution \cite{S--Foster11}.
The fractionalization waves shown in Fig.~2 were obtained with $\zeta = 0$. Below we show in Fig.~\ref{SFig--0.7,Reg}
that retaining $\zeta > 0$ 
does not qualitatively affect the fractionalized case, except at long times when it regularizes the growth of the ``supersolitons.''
This is consistent with previous results for an \emph{XXZ} chain lattice quench in 1+1-D \cite{S--Foster11}. 

For the numerical evaluation of Eq.~(4) in the main text, we use a Gaussian potential as in Eq.~(6).
In fact, since the linearized theory in Eq.~(2) keeps momentum modes with $|k_y| \leq \pi/(2 a)$, 
we retain lattice-scale resolution in the $y$-direction. To be precise, we can resolve \emph{pairs} of 
chains in the initial state; the members of a pair are in fact encoded in the $\tau$-pseudospin degree of freedom
of $\psi_{\sigma,\tau}(x,y)$. 
Therefore we write 
\begin{align}\label{S--PhiComp}
\begin{gathered}
	\kappa \Phi(x,y) 
	\equiv
	Q \, \Phi_x(x) \, \Phi_y(y = 2 a j),
\\
	\Phi_x(x) 
	=
	\frac{1}{\sqrt{\pi}\Delta_x}
	\,
	\exp\left({-x^2}/{\Delta_x^2}\right),
\quad
	\Phi_y(y = 2aj) 
	= 
	N(a,\Delta_y)
	\,
	\exp\left[{-(2 a j)^2}/{\Delta_y^2}\right].
\end{gathered}
\end{align}
Here $j \in \mathbb{Z}$ indexes chain pairs. 
The normalization factor is 
\[
	N(a,\Delta_y) 
	= 
	\left[{2a}\,{\vartheta_3\left(0,e^{-4a^2/\Delta_y^2}\right)}\right]^{-1} 
	\rightarrow 
	\frac{1}{\sqrt{\pi}\Delta_y} \text{as}\ a \rightarrow 0.
\]
In this equation $\vartheta_3(u,q)$ is the Jacobi theta function.


\section{IV.\ Quench dynamics}

The explicit solution to the density $\rho(t,x,y)$ obtains from Eq.~(4), 
using Eqs.~(5), (\ref{S--G2+1}), (\ref{S--G1+1Tot}) and (\ref{S--PhiComp}).
It can be written as follows:
\begin{align}\label{S--rho(t)Evol}
	\rho(t,x,y) 
	= 
	-
	\frac{Q}{2}
	\,
	\Phi_x(x-t)
	\,
	\Phi_y(y)
	+ 
	c_\eta \, Q
	\left[\mathcal{I}_1(t,x,y) + \mathcal{I}_2(t,x,y)\right]
	+
	\left(x \Rightarrow -x\right),
\end{align}
where 
\begin{align}\label{S--I_1}
	\mathcal{I}_1(t,x,y) 
	= 
	2 a
	\int_{-t}^{t}
	d x_2 
	\int_{-\frac{\pi}{2 a}}^{\frac{\pi}{2 a}}
	\frac{d k_y}{2 \pi} 
	\,
	G^\pup{1}\left[t,x_2;m(k_y)\right]		
	\left[\frac{\alpha^2}{(t - x_2)^2 + \zeta^2}\right]^{\eta/2}
	\frac{
	\int_{x - t}^{x - x_2} 
	d z
	\, 
	\Phi_x(z)
	\,
	\Phi_y(y)}{
	(x_2 - t)
	},
\end{align}
and
\begin{align}\label{S--I_2}
	\mathcal{I}_2(t,x,y) 
	=&\,
	a
	\int_{-t}^t 
	d x_1 
	\int_{-t}^{t}
	d x_2 
	\int_{-\frac{\pi}{2 a}}^{\frac{\pi}{2 a}} 
	\frac{d k_y}{2 \pi} 
	\int_{-\frac{\pi}{2 a}}^{\frac{\pi}{2 a}} 
	\frac{d k_y'}{2 \pi} 
	\left\{
	\begin{aligned}
		&\,
		G^\pup{1}\left[t,x_1;m(k_y)\right] G^\pup{1}\left[t,x_2;m(k_y')\right]
		\\&\,
		+
		G^\pup{2}\left[t,x_1;m(k_y)\right] G^\pup{2}\left[t,x_2;m(k_y')\right]
	\end{aligned}
	\right\}
	\nonumber\\
	&\,
	\times
	\cos\left[(k_y' - k_y) y\right]
	\,
	\tilde{\Phi}_y\left(k_y' - k_y \right)
	\left[\frac{\alpha^2}{(x_2 - x_1)^2 + \zeta^2}\right]^{\eta/2}
	\frac{\int_{x - x_1}^{x - x_2} d z \,\Phi_x(z)}{(x_2 - x_1)}.
\end{align}
The transform of $\Phi_y(y)$ is defined via 
\begin{equation}\label{S--InitialTheta}
	\tilde{\Phi}_y(p) 
	\equiv
	\frac{1}{\vartheta_3\left(0,e^{-4 a^2/\Delta_y^2}\right)}
	\sum_j 
	e^{-i p (2 a j)}
	\exp\left[\frac{-4(aj)^2}{\Delta_y^2}\right]
	= 
	\frac{\vartheta_3\left(a p; e^{-4a^2/\Delta^2_y}\right)}{\vartheta_3\left(0,e^{-4a^2/\Delta_y^2}\right)}.
\end{equation}
We evaluate the integrals in Eqs.~(\ref{S--I_1}) and (\ref{S--I_2}) numerically to compute the density profiles shown in 
Figs.~2 and 3 of the main text.
We provide two independent checks for our numerical integration routine in
Figs.~\ref{SFig--EPlots} and \ref{SFig--1D}, discussed below.


\section{V.\ Single-particle wavepacket dynamics in the lattice-regularized Dirac semimetal}

\begin{figure}[b!]
    \centering
    \begin{minipage}{0.43\textwidth}
        \centering
        \includegraphics[width=0.9\textwidth]{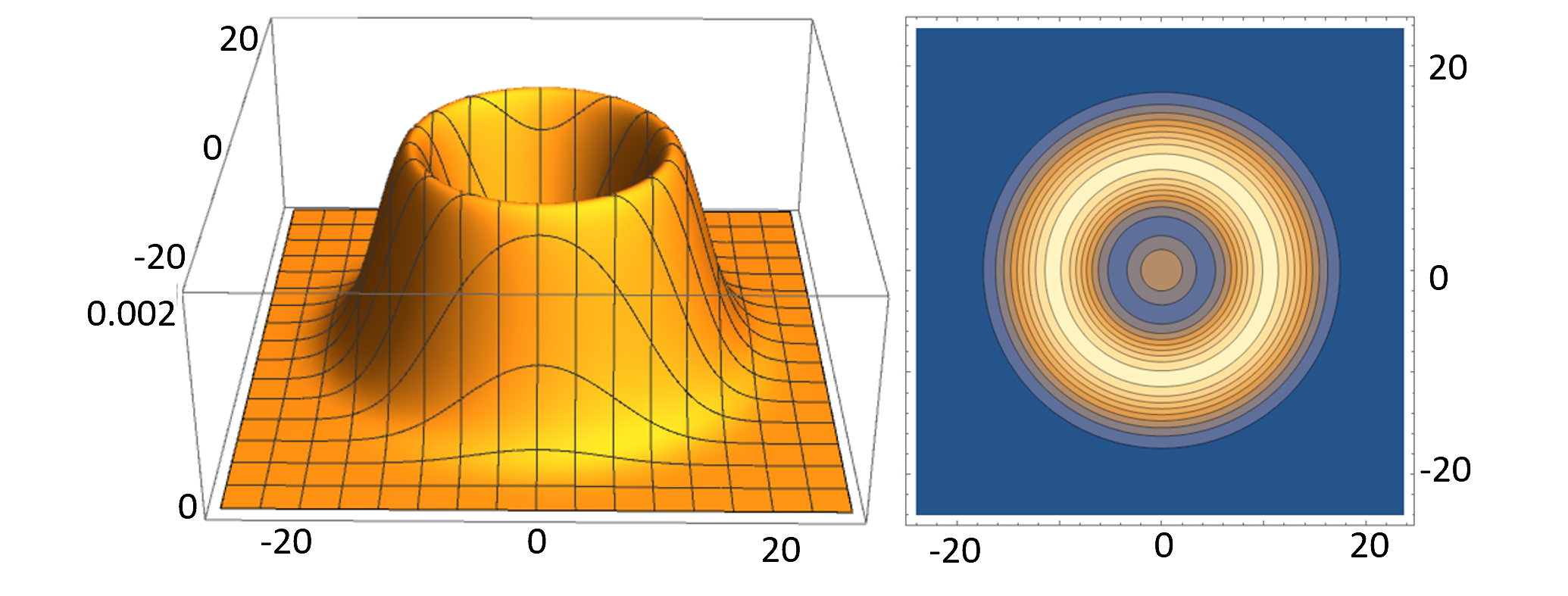}
        \includegraphics[width=0.9\textwidth]{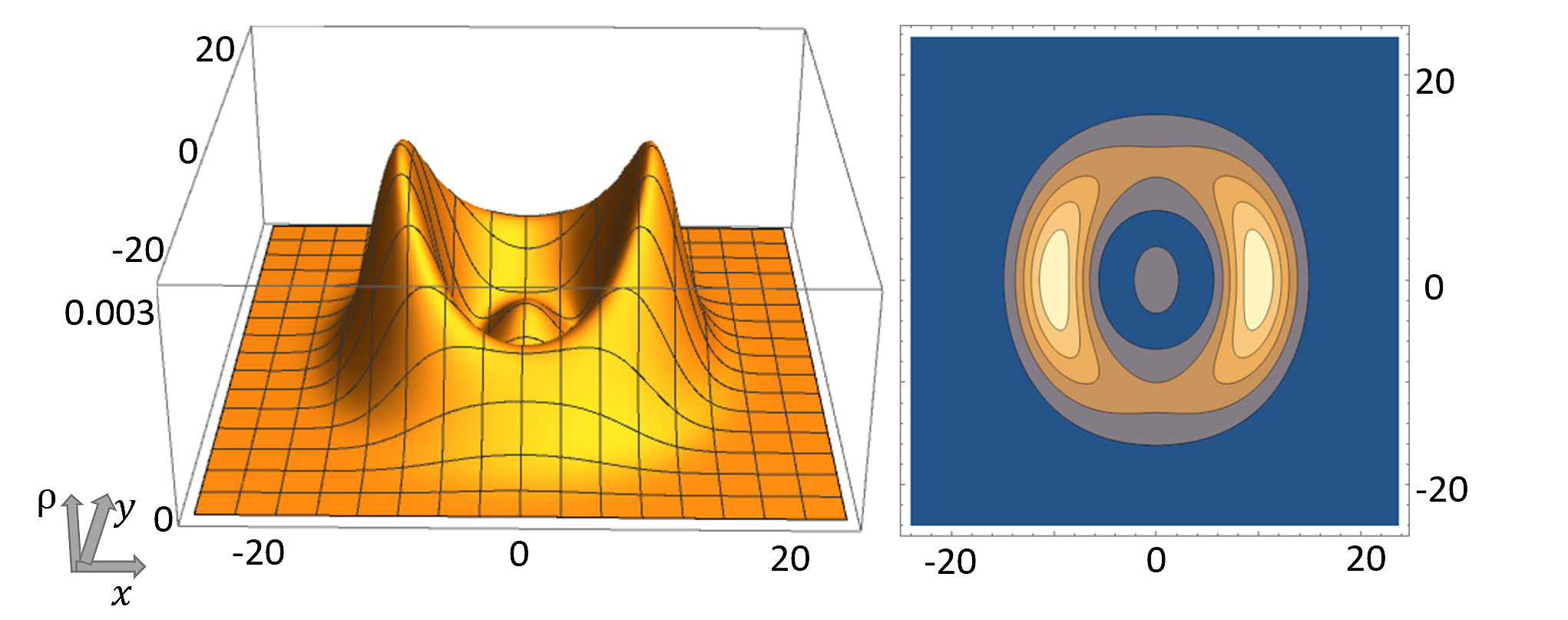}
    \end{minipage}
	\hspace{14pt}
        \begin{minipage}{0.43\textwidth}
        \centering
        \includegraphics[width=0.9\textwidth]{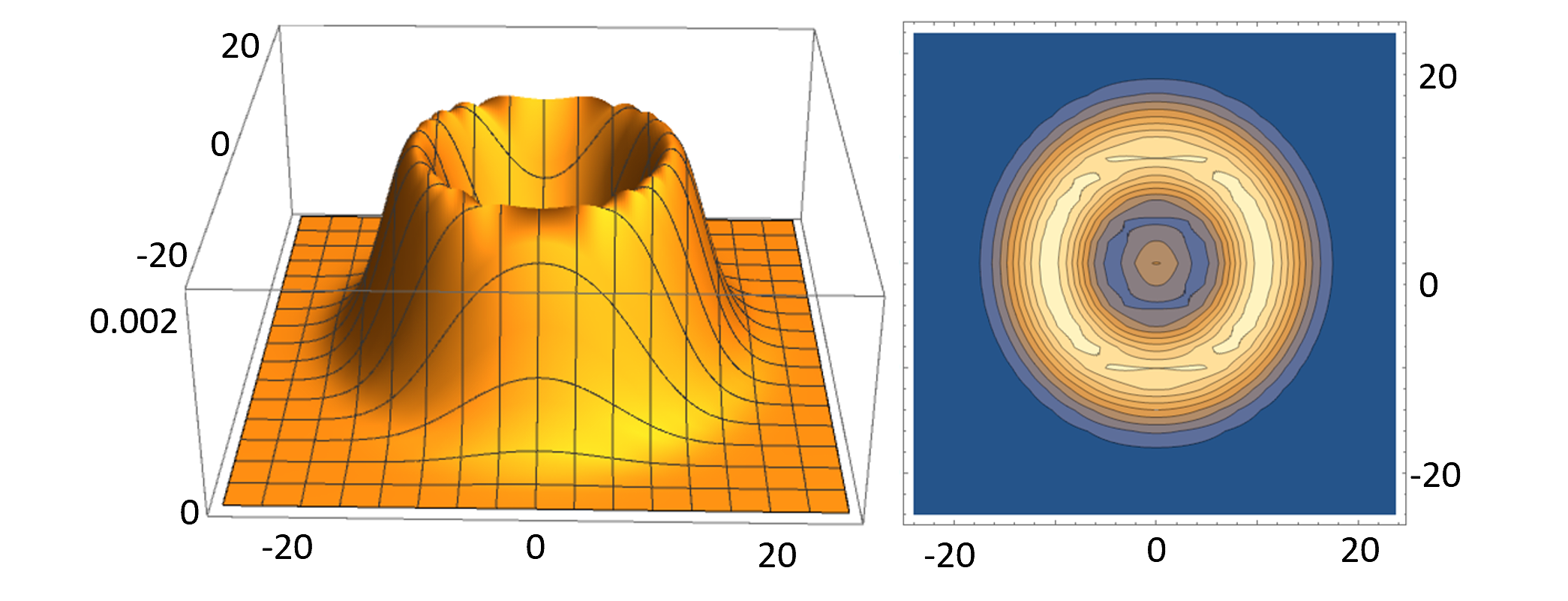}
        \includegraphics[width=0.87\textwidth]{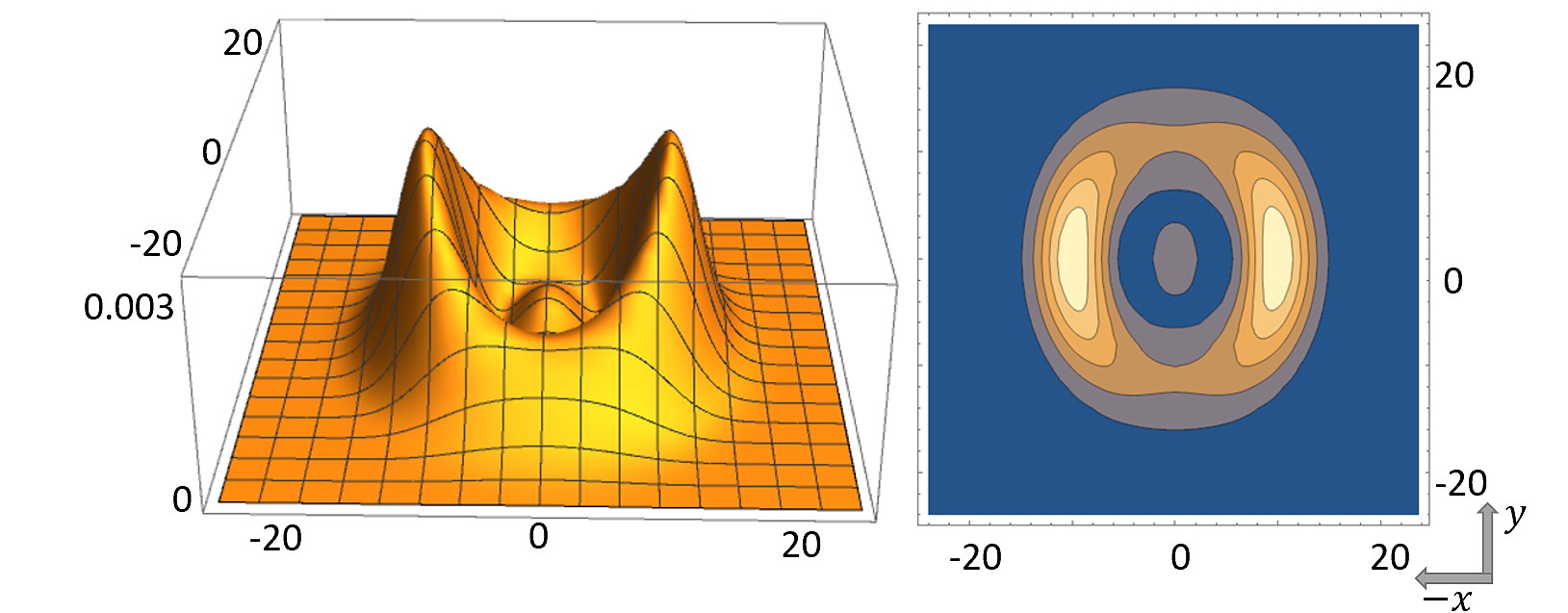}
    \end{minipage}\hfill
    \caption{	Time-evolved probability profile for an initial Gaussian wavepacket 
		[Eq.~(\ref{S--PsiGauss})]
		for a pure massless Dirac band structure (left) 
		and the lattice-regularized evolution operator in Eq.~(\ref{S--G2+1})
		(right). 
		The top pair of images corresponds 
		to an isotropic initial condition with $\Delta_x = \Delta_y = 2.5a$, while the bottom 
		pair of images corresponds to an initial condition with $\Delta_x = 2a, \Delta_y = 3a$, 
		as in the many-body quench Figs.~2 and 3 in the main text. The images are obtained at time $t = 9$,
		with $v_F = 1$.}
	\label{SFig--QM}
\end{figure}

Eq.~(\ref{S--psiEvol}) also describes the evolution of a wavepacket in one-particle quantum mechanics,
provided that $\psi(t,x,y)$ is re-interpreted as a Dirac (spinor) wave function, with
$\psi(x,y)$ the initial condition. 
We consider a Gaussian initial wavepacket with zero average probability current 
in the $x$- and $y$-directions:
\begin{align}\label{S--PsiGauss}
	\psi(x,y)
	=
	\frac{1}{\sqrt{\pi \Delta_x \Delta_y}}
	\exp\left(- \frac{x^2}{2 \Delta_x^2} - \frac{y^2}{2 \Delta_y^2}\right)
	\ket{\uparrow_x}, 	
\end{align}
where $\sigh^1 \ket{\uparrow_x} = \ket{\uparrow_x}$. 
Using the Green's function in Eq.~(\ref{S--G2+1}), we time-evolve from the initial condition in Eq.~(\ref{S--PsiGauss}).
The resulting probability density $\psi^\dagger\psi(t,x,y)$ is shown in Fig.~\ref{SFig--QM}. 
To demonstrate that the lattice regularization in the $y$-direction does not significantly affect the dynamics on the time scales 
we are interested in, we compare the profiles computed using $\hat{G}(t,x,y)$ in Eq.~(\ref{S--G2+1}) 
to that obtained from a massless, isotropic 2D Dirac Hamiltonian with purely linear dispersion.

The single-particle quantum evolution shown in Fig.~\ref{SFig--QM} exhibits 
light-cone propagation of probability, with a circular front for an isotropic initial condition.  
The key observation is that this bears no resemblance to the density dynamics for the many-particle quench. 
This is true for the interacting quench with $\eta > 0$, which produces supersolitons that propagate
relativistically along the chains (Fig.~2), 
as well as the \emph{noninteracting} quench with $\eta = 0$ (Fig.~3), 
which produces dispersive transport perpendicular to the chains.

The deviation of the $\eta = 0$ quench from one-particle quantum mechanics depicted in Fig.~\ref{SFig--QM} 
is due to Pauli blocking.
As explained in the main text, Pauli blocking 
suppresses large $v_x$ in the Wigner velocity distribution for the noninteracting quench, 
shown in the bottom panel of Fig.~4.
The latter plot also
demonstrates that the noninteracting quench excites particles with a broad range of $v_y$. 
The $y$-oriented dispersive propagation shown in Fig.~3 therefore
results from the excitation of modes at all $k_y$ along the ``vertical strips'' retained in the reduced Brillouin zone
[Eq.~(2)]. 

\begin{figure}[h!]
    \centering
    \includegraphics[width=0.9\textwidth]{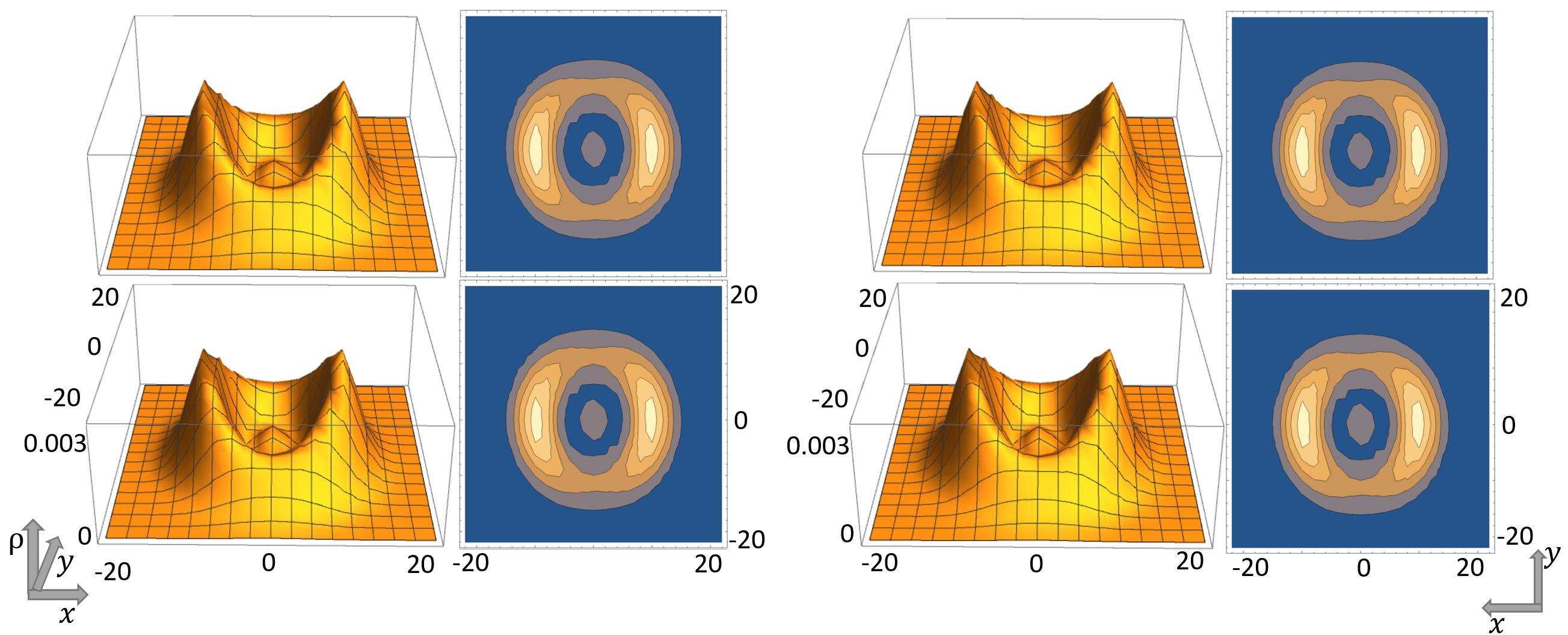}
    \hfill
    \caption{	Time-evolved probability profiles for an initial Gaussian wavepacket, carried out in momentum space to include full lattice detail. 
		The top row uses the continuum approximation in the $x$-direction $m(k_x) \rightarrow k_x$ in Eq.~(\ref{S--GFull}),
		while the bottom row employs the lattice function $m(k_x)$. 
		The left column uses the continuum approximation in the $y$-direction $m(k_y) \rightarrow k_y$,
		while the right column employs the lattice function $m(k_y)$
		All images correspond to an initial condition with $\Delta_x = 2a, \Delta_y = 3a$, 
		as in the many-body quench Figs.~2 and 3 in the main text. The images are obtained at time $t = 9$,
		with $v_F = 1$.
		We see that the top two images here successfully recreate the bottom two images in \ref{SFig--QM}, 
		and that the images here are essentially indistinguishable.}
	\label{SFig--QMLattice}
\end{figure}

We also check that keeping full lattice-scale detail in the $x$-direction does not meaningfully alter the dynamics. 
We use the momentum-space form of the full-lattice Green's function [Eq.~(\ref{S--GFull})] and the discretized 
Gaussian initial condition [Eq.~(\ref{S--InitialTheta})] to carry out the time-evolution convolution in momentum space. 
The results are given in Fig.~\ref{SFig--QMLattice}, where all combinations of continuum and lattice propagator forms are 
compared and found to be essentially identical. 
This further supports our use of the $x$-continuum version of the Green's function for time evolution.

\section{VI.\ Additional fractionalization wave numerics}

In this section we exhibit additional numerical results for the 
1D-to-2D quench dynamics described in the main text.

\subsection{A.\ Results for $\eta = 0.2$}

\begin{figure}[t!]
    \centering
    \begin{minipage}{0.4\textwidth}
        \centering
        \includegraphics[width=0.9\textwidth]{fig3.png}
       \caption{Time evolution of the density $\rho(t,x,y)$ with $\eta = 0$ (same as Fig.~3).} 
       \label{SFig--3Redux}
    \end{minipage}\hspace{14pt}
        \begin{minipage}{0.41\textwidth}
        \centering
        \includegraphics[width=0.9\textwidth]{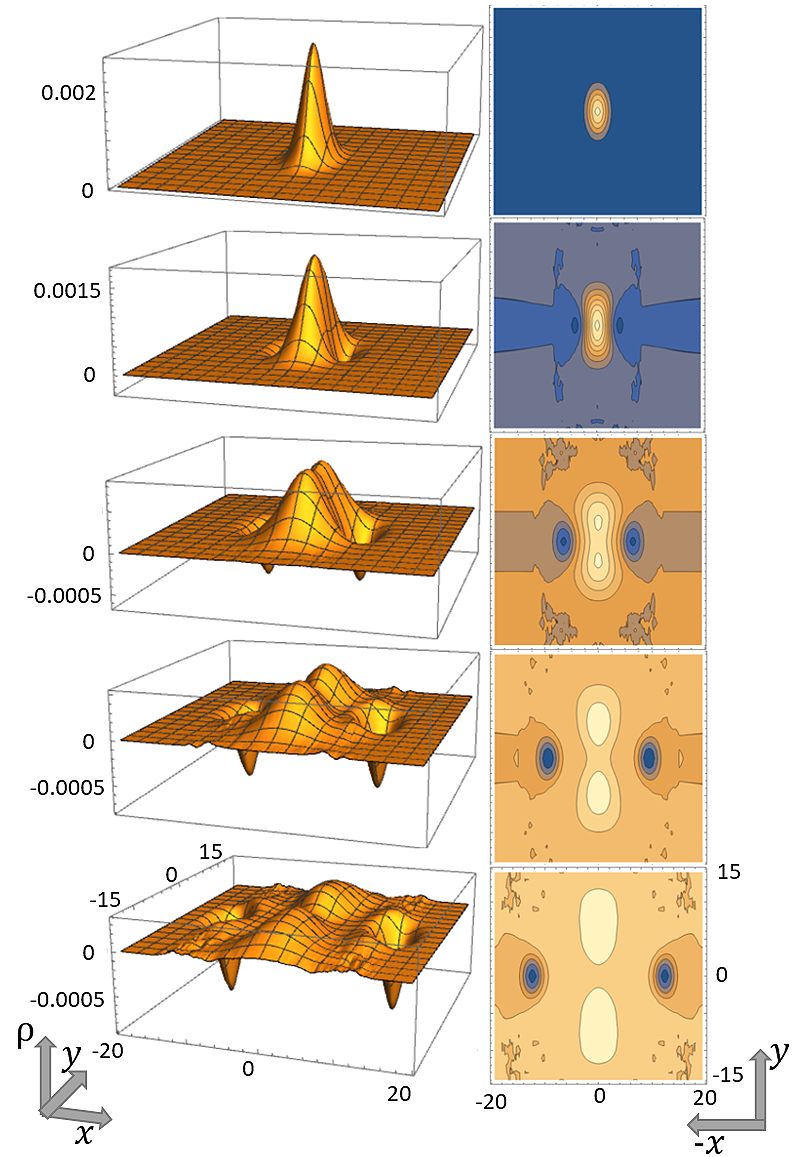}
       \caption{Time evolution of the density $\rho(t,x,y)$ as in Fig.~2, but with $\eta = 0.2$.} 
	\label{SFig--0.2}
    \end{minipage}\hfill
\end{figure}

Fractionalization waves as in Fig.~2 are shown in Fig.~\ref{SFig--0.2}.
All parameters are identical to that in Fig.~2, except that the
initial-state-fermion anomalous dimension has been reduced to $\eta = 0.2$ in Fig.~\ref{SFig--0.2}.
The result is shown alongside the noninteracting $\eta = 0$ case for comparison.

\subsection{B.\ UV regularization}

The results shown in Figs.~2 and \ref{SFig--0.2} were obtained using the initial
state correlator in Eq.~(5), with the parameter $\zeta = 0$. 
Nonzero $\zeta$ can arise due to the influence of irrelevant operators \cite{S--Foster11}.
In Fig.~\ref{SFig--0.7,Reg}, we plot the same evolution shown in Fig.~2 of the main text,
except that here $\zeta = 1$ (in units of the lattice spacing $a$). 
This is plotted alongside the $\zeta = 0$ result from Fig.~2 for comparison. 

The main effect of nonzero $\zeta$ is to amputate the power-law growth of the 
supersolitons at large times \cite{S--Foster11}, as shown in Fig.~\ref{SFig--0.7,Reg}.
A quantitative estimate for $\zeta$ would require knowledge of the 
microscopic details for the pi-flux lattice quench implementation. Regardless,
the primary effect described in the main paper for the case of a fractionalized
initial condition is unchanged, i.e.\ the transient generation of directed, 
relativistically propagating waves along the wires.

\begin{figure}[t!]
    \centering
    \begin{minipage}{0.41\textwidth}
        \centering
        \includegraphics[width=0.9\textwidth]{fig2.png}
 \caption{Time evolution with $\eta = 0.7$ (the same as Fig.~2 in the main text).} 
	\label{SFig--2Redux}
    \end{minipage}\hspace{14pt}
        \begin{minipage}{0.41\textwidth}
        \centering
        \includegraphics[width=0.9\textwidth]{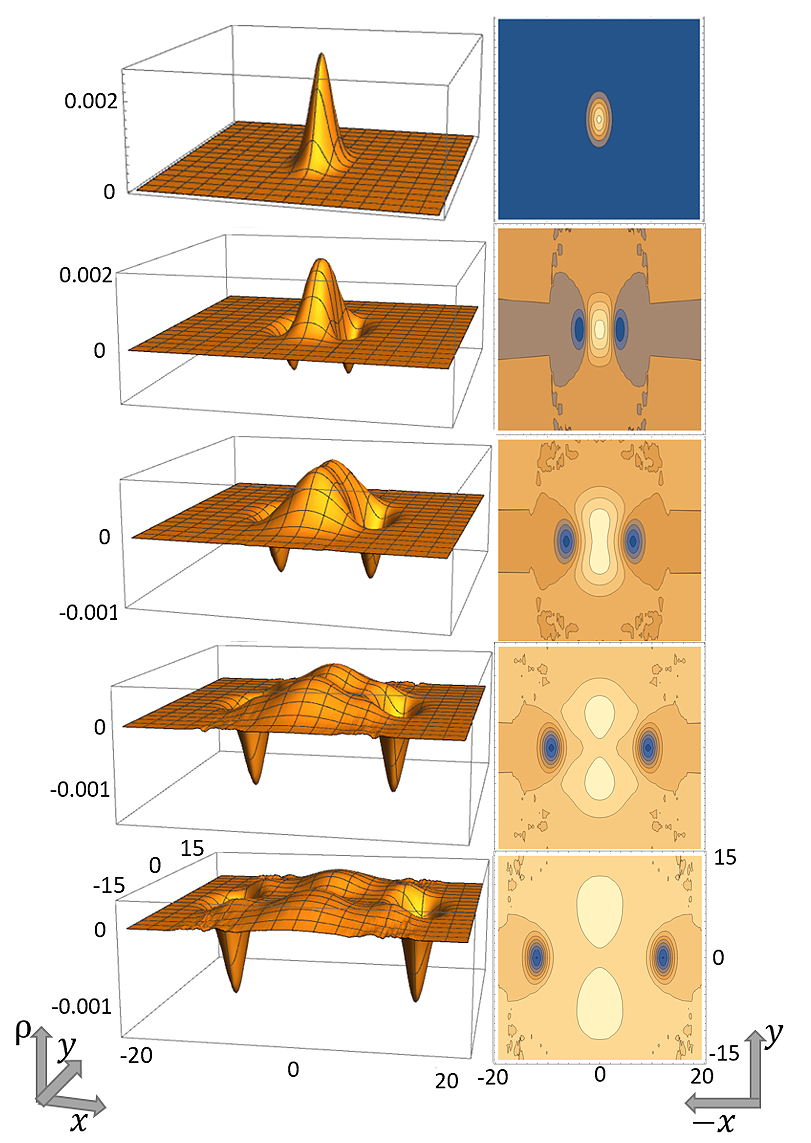}
         \caption{Time evolution with $\eta = 0.7$ as in Figs.~2 and \ref{SFig--2Redux}, 
	except that here the UV regularization parameter $\zeta = a = 1$ [see Eq.~(5)],
	while we have taken $\zeta = 0$ in Figs.~2, \ref{SFig--2Redux} and in Fig.~\ref{SFig--0.2}.
	The main effect of nonzero $\zeta$ is the amputation of the supersoliton growth at large times.}
	\label{SFig--0.7,Reg}
    \end{minipage}\hfill
\end{figure}

\subsection{C.\ Error control in the numerics}

The time evolution of our quench dynamics should conserve particle number. 
Since the entire weight of the (right-moving piece of the) density bump is encoded in the first term
on the right-hand-side of Eq.~(\ref{S--rho(t)Evol}), it must be that the integrated weight
$\mathcal{N}_1(t) + \mathcal{N}_2(t) = 0$ at all times. 
Here 
\begin{align}
	\mathcal{N}_{1,2}(t) \equiv \int d x \, d y \, \mathcal{I}_{1,2}(t,x,y),
\end{align}
where $\mathcal{I}_{1,2}(t,x,y)$ were defined in Eqs.~(\ref{S--I_1}) and (\ref{S--I_2}). 

\begin{figure}[b!]
\centering
\includegraphics[angle=0,width=.4\textwidth]{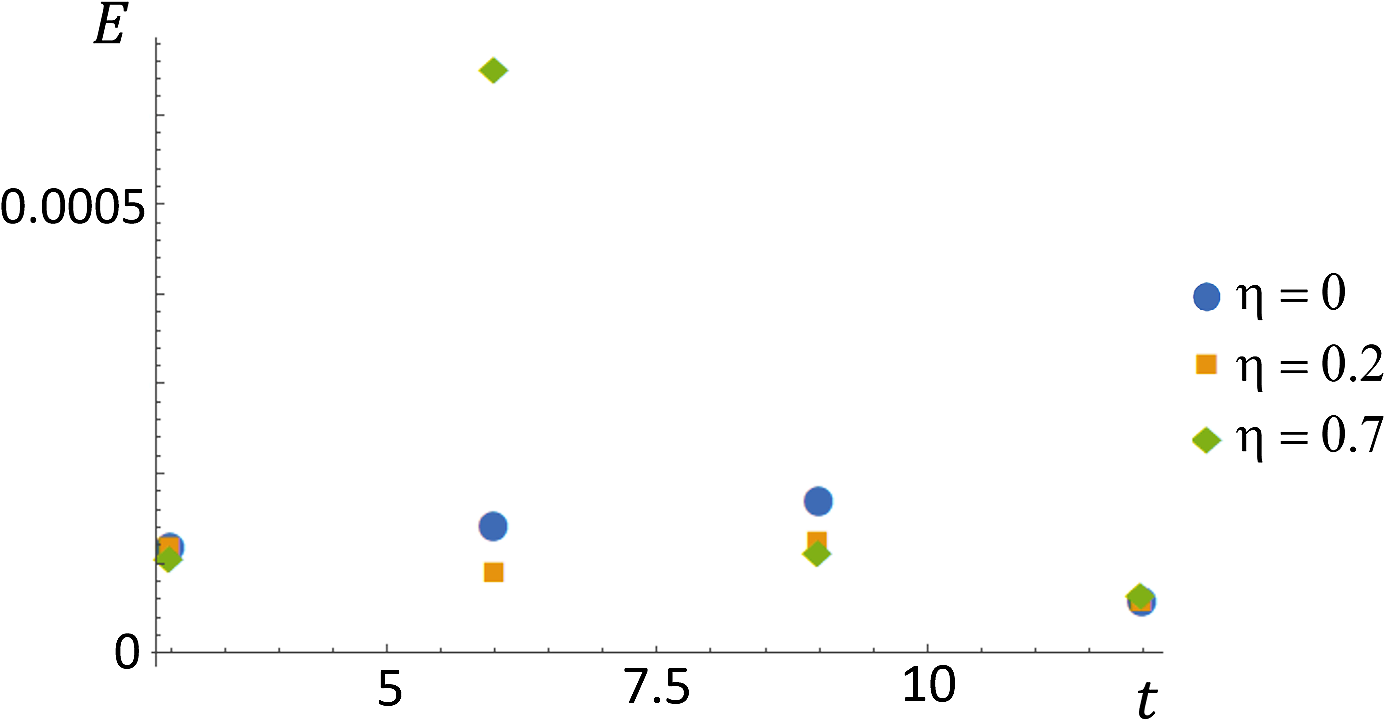}
\caption{Error ratio $E(t)$ defined by Eq.~(\ref{S--Error}) for the 
time-slice profiles given for the various many-body quenches in 
Figs.~3, \ref{SFig--3Redux} ($\eta = 0$), 
Fig.~\ref{SFig--0.2} ($\eta = 0.2$),
and
Figs.~2, \ref{SFig--2Redux} ($\eta = 0.7$).}
\label{SFig--EPlots}
\end{figure}

We define the error ratio 
\begin{align}\label{S--Error}
	E(t)
	\equiv
	\frac{|\mathcal{N}_1+\mathcal{N}_2|}{|\mathcal{N}_1|+|\mathcal{N}_2|}.
\end{align}
This quantity is plotted in Fig.~\ref{SFig--EPlots} for $\eta \in \{0,0.2,0.7\}$ and $\zeta = 0$.

\subsection{D.\ Full Lattice Quench, Noninteracting case.}

As a further check on our results and on our use of the continuum approximation in the $x$-direction, 
we re-calculate the many-body quench dynamics in the noninteracting case by performing an exact position-space diagonalization
of the 2D lattice model. We first find the exact eigenstates of the single-particle Hamiltonian and construct an initial many-body 
ground state (at half-filling) as a Slater determinate of the lower half of the eigenstates. 
We then calculate all the eigenstates of the post-quench Hamiltonian and use these to time-evolve the initial state.

Our results are plotted below in Figs.~\ref{SFig--22Lat} and \ref{SFig--33Lat} for $66$-by-$66$ grids of lattice sites with periodic boundary conditions. 
The density bump disperses along the $y$-axis, qualitatively in full agreement with our earlier calculation shown in Fig.~3. 
For the same size bump used in the main text, we do indeed see slight deviations 
in Fig.~\ref{SFig--22Lat}
from the profiles produced in Fig.~3, 
due to some combination of finite-size lattice effects and details neglected from the continuum approximation of the initial state correlation function, Eq.~(\ref{S--IC-Corr}). 
However, as we increase the size of the initial wavepacket relative to the lattice spacing
as in Fig.~\ref{SFig--33Lat}, the exact results begin to strongly resemble the earlier ones produced by 
the approximate initial correlator in Eq.~(5). The results here indicate that lattice-scale details neglected in our earlier calculation take the 
form of small ripples on the wavefront, but that these can be removed by choosing a larger initial bump. 
The small bump size used in the figures of the main text
was chosen to conserve computational resources while demonstrating the essential physics.

\begin{figure}[t!]
    \centering
    \begin{minipage}{0.41\textwidth}
        \centering
        \includegraphics[width=0.9\textwidth]{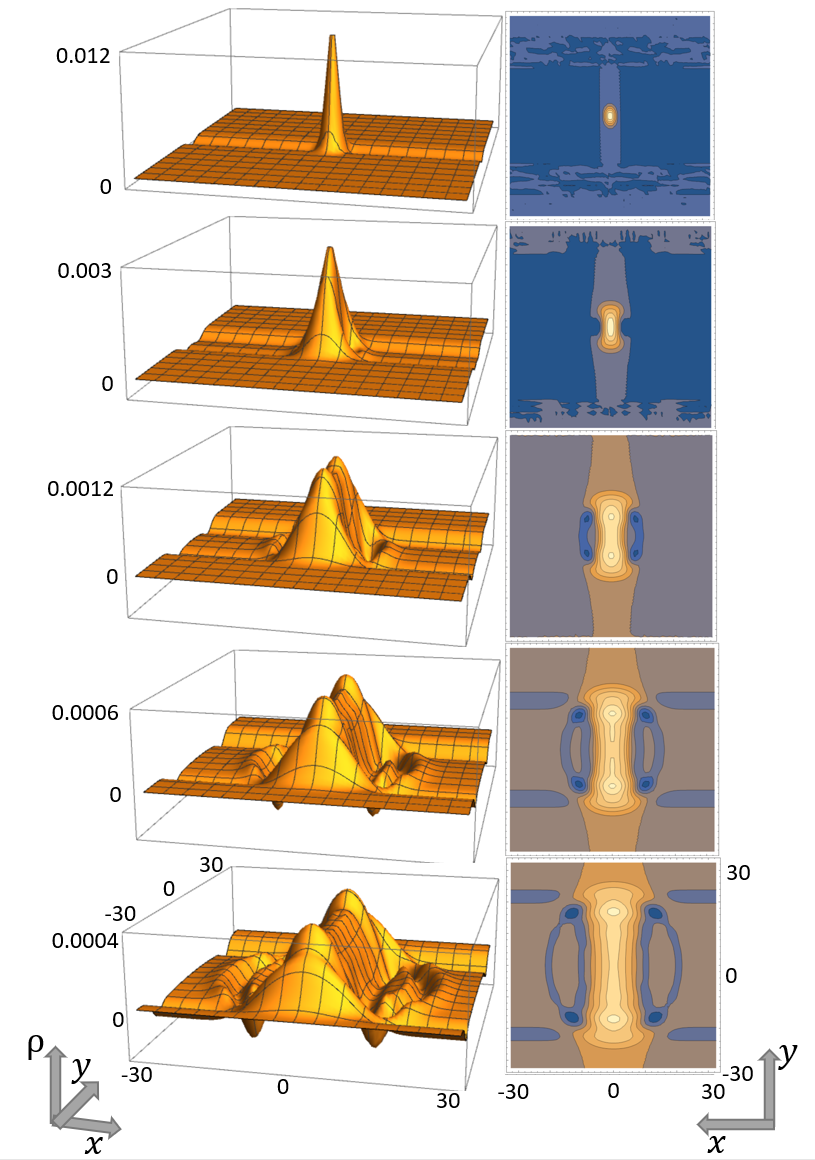}
 \caption{Time evolution with $\eta = 0$ using exact diagonalization of 66-by-66 lattice Hamiltonian. 
	The initial wavepacket widths are $\Delta_x = 2a, \Delta_y = 3a$, as in Fig.~3 of the main text.}
	\label{SFig--22Lat}
    \end{minipage}\hspace{14pt}
        \begin{minipage}{0.41\textwidth}
        \centering
        \includegraphics[width=0.95\textwidth]{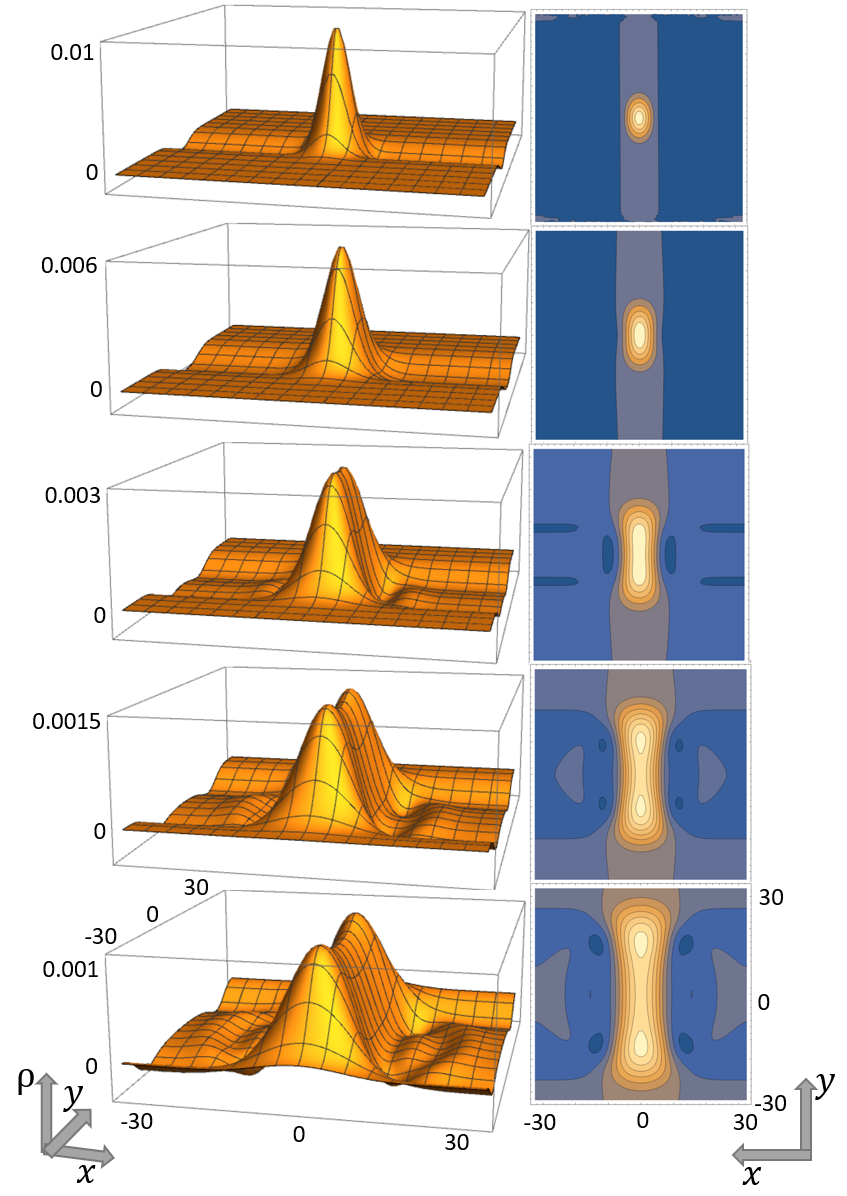}
         \caption{Time evolution with $\eta = 0$ using exact diagonalization of 66-by-66 lattice Hamiltonian. 
	The initial wavepacket widths are $\Delta_x = 4a, \Delta_y = 6a$, twice as large as in Fig.~3 of the main text.}
	\label{SFig--33Lat}
    \end{minipage}\hfill
\end{figure}


\section{VII.\ 1D initial condition: numerics vs.\ asymptotic analysis}

\begin{figure}[t!]
\centering
\includegraphics[angle=0,width=.65\textwidth]{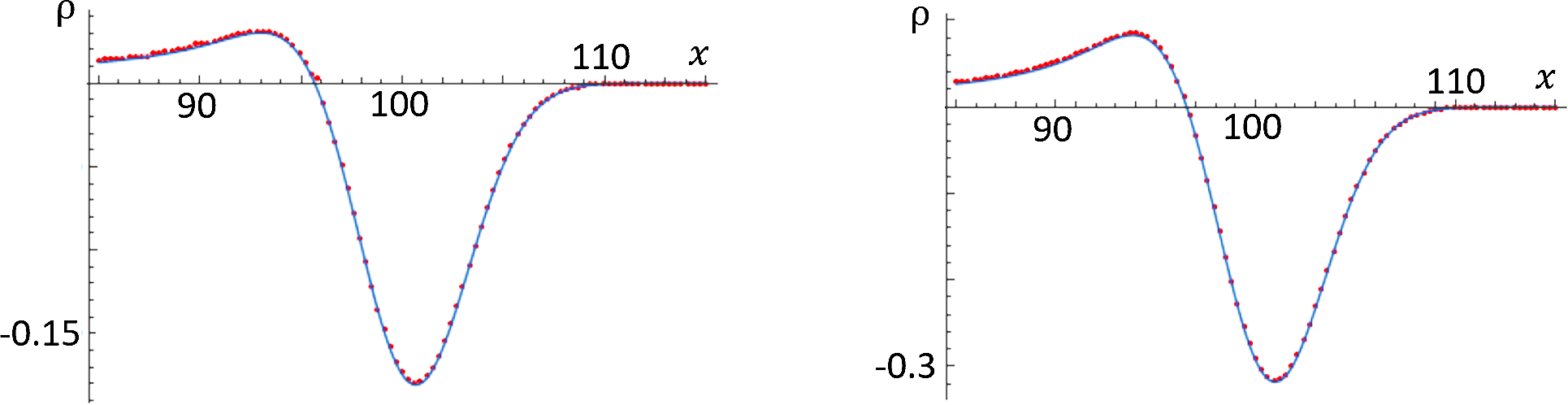}
\caption{Comparison of the numerically-calculated supersoliton profile (red dots)
to the analytic asymptotic approximation
in Eq.~(\ref{S--Asym}) (blue solid lines). 
The profile is shown at time $t = 100$, with initial parameters $\Delta_x = 6$, 
$\eta = 0.4$ (left), 
and 
$\eta = 0.6$ (right).}
\label{SFig--1D}
\end{figure}

The integrals in Eqs.~(\ref{S--I_1}) and (\ref{S--I_2}) evaluated at large times can be scrutinized via asymptotic analysis \cite{S--Foster11}. 
For the two-dimensional initial condition in Eq.~(6) this is rather complicated. 
A simpler case takes the initial density inhomogeneity to be uniform in the $y$-direction,
$\Phi_y(y) = 1$ [$\tilde{\Phi}_y(k_y) = 2 \pi \delta(k_y)$]. 
In this case asymptotic analysis performed 
on 
Eqs.~(\ref{S--I_1}) and (\ref{S--I_2})
along the lines of Appendix A in Ref.~\cite{S--Foster11} leads to 
\begin{align}\label{S--Asym}
	\rho(t,x) 
	= 
	-
	\frac{Q}{2}\Phi(x-t) 
	- 
	\frac{Q}{\Delta}
	\left(\frac{\alpha^2 t}{\sqrt{2}\Delta}\right)^{\eta/2}
	\frac{1}{\sqrt{\pi}}
	\left(\frac{2}{a}\right)^{\eta }
	\frac{\Gamma[1 - \eta]}{\Gamma[1+\eta/2]}
	e^{-(t-x)^2/2\Delta^2}
	\mathcal{D}_{\eta/2}\left[\sqrt{2}\left(\frac{t - x}{\Delta}\right)^2\right]
	+
	\left(x \Rightarrow -x\right).
\end{align}
Here $D_\nu(x)$ denotes the parabolic cylinder function. 
In Fig.~\ref{SFig--1D}, we compare 
the result in Eq.~(\ref{S--Asym}) to numerical
integration for different values of $\eta$. Beyond the error analysis
presented in Fig.~\ref{SFig--EPlots} for the 2D initial condition, this is a second
check on the numerical integration routine used to obtain Figs.~2 and 3 
in the main text from Eqs.~(\ref{S--I_1}) and (\ref{S--I_2}) 
(since the same routine is employed for the 1D initial condition and
the results shown in Fig.~\ref{SFig--1D}).

\section{VIII.\ Calculation of the Wigner distribution}

The post-quench Hamiltonian in Eq.~(2) 
(with vanishing external potential $\Phi = 0$, neglecting interactions in $H_I$)
can be diagonalized in terms of canonically-quantized creation and annihilation operators as 
\begin{equation}\label{S--HDecomp}
	H 
	= 
	v_F 
	\int_{-\Lambda}^\Lambda
	\frac{d k_x}{2 \pi}
	\int_{-\frac{\pi}{2a}}^{\frac{\pi}{2a}}
	\frac{d k_y}{2 \pi}
	\,
	\e_{\vex{k}} 
	\left[
		a^\dagger(\vex{k}) \, a(\vex{k}) 
		+ 
		b^\dagger(\vex{k}) \, b(\vex{k}) 
	\right],
\end{equation}
where
\begin{equation}\label{S--eDef}
	\e_{\vex{k}} 
	= 
	\sqrt{k_x^2 + m^2(k_y)}.
\end{equation}
Here $a(\vex{k})$ [$b(\vex{k})$] annihilates a conduction band particle (valence band hole) with momentum $\vex{k}$, 
we will set $v_F = 1$, and $m(k_y)$ is defined by Eq.~(3). 
Since Eq.~(2) is invariant under $\tau$-space SU(2) rotations, we suppress $\tau$-pseudospin indices
throughout this section. 

The Dirac spinor $\psi \rightarrow \psi_{\sigma}$ 
is expressed in terms of the creation and annihilation operators via
\begin{equation}\label{S--psi_Decomp}
	\begin{bmatrix}
	\psi_1(\vex{k})\\
	\psi_2(\vex{k})
	\end{bmatrix}
	= 
	\frac{1}{\sqrt{1+s^2(\vex{k})}}
	\begin{bmatrix}
	1\\
	i s(\vex{k})
	\end{bmatrix}
	a(\vex{k})
	+
	\frac{1}{\sqrt{1+s^2(-\vex{k})}}
	\begin{bmatrix}
	1\\
	-i s(-\vex{k})
	\end{bmatrix}
	b^\dagger(-\vex{k}),
\end{equation}
where
\begin{equation}
	s(\vex{k}) 
	= 
	\frac{\e_{\vex{k}} - k_x}{\left|m(k_y)\right|}.
\end{equation}

The band velocities and momenta are related by 
\begin{align}
	v_x 
	= 
	\frac{k_x}{\e_{\vex{k}}},
\quad
	v_y 
	= 
	\frac{b \, m(k_y)}{\e_{\vex{k}}}
	\cos(k_y a),
\end{align}
\begin{align}\label{S--kx,ky}
	k_x 
	= 
	\frac{m(v_x,v_y) \, v_x}{\sqrt{1-v_x^2}},
\quad
	k_y
	= 
	\frac{1}{a}
	\arcsin\left[
	\frac{a}{b}
	m(v_x,v_y)
	\right].
\end{align}
Here
\begin{align}
	m(v_x,v_y) 
	=
	\frac{\sgn(v_y)}{a}
	\sqrt{b^2 - \left(\frac{v_y^2}{1-v_x^2}\right)}
\end{align}
is the mass parameter $m(k_y)$ [Eq.~(3)] rewritten in terms of velocities. 
The velocities are constrained to the (elliptical) disk in Eq.~(8). 
The Jacobian corresponding to the change of variables 
$(k_x,k_y) \Rightarrow (v_x,v_y)$ 
is 
\begin{equation}\label{S--Jacob}
	\mathcal{J}(v_x,v_y)
	\equiv
	\left|\frac{\partial(k_x,k_y)}{\partial(v_x,v_y)}\right| 
	= 
	\frac{1}{a^2}
	\frac{1}{(1 - v_x^2)^2}.
\end{equation}

As in \cite{S--Foster11}, 
we define the ground-state Wigner function for the right-moving fermion $\psi_1(x) \equiv R(x)$ as 
\begin{equation}\label{S--RightWigner}
	\delta n_R(k_x,k_y;R_x,R_y) 
	\equiv
	\int 
	d x_d 
	\,
	d y_d 
	\,
	e^{- i k_x x_d - i k_y y_d}
	\left\langle
	R^\dagger\left(R_x - \frac{x_d}{2},R_y - \frac{y_d}{2}\right)
	\,
	R\left(R_x + \frac{x_d}{2},R_y + \frac{y_d}{2}\right)
	\right\rangle_{\Phi},
\end{equation}
where the subscript $\Phi$ denotes the linear response to the external potential $\Phi(x,y)$ in Eq.~(2a).  
The initial state correlation function for the pre-quench ground state formed from 
uncoupled chains is 
\begin{equation}
	\left\langle
	R^\dagger\left(R_x - \frac{x_d}{2},R_y - \frac{y_d}{2}\right)
	\,
	R\left(R_x + \frac{x_d}{2},R_y + \frac{y_d}{2}\right)
	\right\rangle_{\Phi}
	=
	\delta(y_d)
	\,
	\mathcal{C}_{\Phi}\left(R_x - \frac{x_d}{2},R_x + \frac{x_d}{2};R_y\right),
\end{equation}
where  
$\mathcal{C}_{\Phi}(x_1,x_2;y)$ is given by Eq.~(5). 
Ignoring for simplicity the ultraviolet regularization parameter $\zeta \equiv 0$, 
Eq.~(\ref{S--RightWigner}) evaluates to \cite{S--Foster11}
\begin{align}\label{S--n_R}
	\delta n_R(k_x,k_y;R_x,R_y) 
	=  
	\frac{c_\eta \alpha^\eta \Gamma(1-\eta) \sin\left(\frac{\pi\eta}{2}\right)}{\eta}
	\Phi_y(R_y)
	\int
	&\,
	\frac{d q}{2\pi}
	\frac{e^{iq R_x}}{q}
	\left[ 
		- Q \, \tilde{\Phi}_x(q)
	\right]
	S(q,k_x,k_y)
\nonumber\\
	\times
	&
	\left\{
		\sgn\left[k_x + \frac{q}{2}\right]\left|k_x + \frac{q}{2}\right|^\eta 
		- 
		\sgn\left[k_x - \frac{q}{2}\right]\left|k_x - \frac{q}{2}\right|^\eta
	\right\},
\end{align}
where we have assumed the separable potential defined by Eqs.~(6) and (\ref{S--PhiComp}).
For the pure right-mover correlator computed in Eq.~(\ref{S--n_R}), the ``structure factor''
$S(q,k_x,k_y) = 1$; therefore $\delta n_R$ is \emph{independent} of $k_y$. It is simply proportional
to the potential profile in the $y$-direction $\Phi_y(R_y)$. 
Eq.~(\ref{S--n_R}) also holds for the prequench Wigner distribution of the left-mover $\psi_2$. 

To compute the Wigner function for the conduction band particle creation and annihilation operators in Eq.~(\ref{S--HDecomp}),
\begin{align}\label{S--n+}
	\delta n_+(k_x,k_y;R_x,R_y)
	\equiv 
	\int 
	\frac{d Q_x \, d Q_y}{(2 \pi)^2} 
	\,
	e^{i Q_x R_x + i Q_y R_y}
	\left\langle
	a^\dagger\left(k_x - \frac{Q_x}{2},k_y - \frac{Q_y}{2}\right)
	\,
	a\left(k_x + \frac{Q_x}{2},k_y + \frac{Q_y}{2}\right)
	\right\rangle_{\Phi},
\end{align}
we exploit the decomposition in Eq.~(\ref{S--psi_Decomp}). 
The result 
for
$\delta n_+(k_x,k_y;R_x,R_y)$
is identical to Eq.~(\ref{S--n_R}), except that now the structure factor
$S(q,k_x,k_y) = B(q,k_x,k_y)$, where
\begin{align}
	B(q,k_x,k_y)
	=
	\beta\left(k_x + \frac{q}{2},k_y\right)
	\beta\left(k_x - \frac{q}{2},k_y\right)
	+
	\beta\left(-k_x + \frac{q}{2},k_y\right)
	\beta\left(-k_x - \frac{q}{2},k_y\right),
\quad
	\beta(k_x,k_y) 
	\equiv
	\sqrt{
		\frac{\e_{\vex{k}} + k_x}{2 \e_{\vex{k}}}
	}.
\end{align}
Although the chiral Wigner function in Eq.~(\ref{S--n_R}) is independent of $k_y$, 
the conduction-band-particle Wigner function defined by Eq.~(\ref{S--n+}) depends on it through 
the structure factor $B(q,k_x,k_y)$. The latter is a function of $k_y$ 
via the band structure $\e_{\vex{k}}$ [Eq.~(\ref{S--eDef})]. 

Using Eqs.~(\ref{S--kx,ky}) and (\ref{S--Jacob}) we can convert to the Wigner velocity distribution defined by Eq.~(7). 
This gives
\begin{align}\label{S--n+_v}
	\delta n_+(v_x,v_y;R_x,R_y) 
	\propto
	\frac{\Phi_y(R_y)}{(1 - v_x^2)^2}
	\int
	&\,
	\frac{d q}{2\pi}
	\frac{e^{iq R_x}}{q}
	\left[ 
		- Q \, \tilde{\Phi}_x(q)
	\right]
	B\left[q,k_x(v_x,v_y),k_y(v_x,v_y)\right]
\nonumber\\
	\times
	&
	\left\{
		\sgn\left[k_x(v_x,v_y) + \frac{q}{2}\right]\left|k_x(v_x,v_y) + \frac{q}{2}\right|^\eta 
		- 
		\sgn\left[k_x(v_x,v_y) - \frac{q}{2}\right]\left|k_x(v_x,v_y) - \frac{q}{2}\right|^\eta
	\right\}.
\end{align}
The plots in Fig.~4 are obtained by numerically integrating Eq.~(\ref{S--n+_v}). 

We can find asymptotic approximations for Eq.~(\ref{S--n+_v}) 
in the limit of large $v_x \rightarrow + 1$ (approaching the maximum band velocity) 
for noninteracting ($\eta = 0$) and interacting ($\eta > 0$) initial conditions.
In the noninteracting case, 
evaluated at $R_x = R_y = 0$ (the center of the pre-quench inhomogeneity),
we obtain
\begin{equation}
	\delta n_+(v_x \rightarrow 1,v_y;0,0)  
	\propto
	\frac{1}{(1 - v_x^2) \left[m(v_x,v_y) \, v_x\right]^{2}}
	\exp\left\{
		-
		\frac{v_x^2 \left[m(v_x,v_y) \, \Delta_x\right]^2}{1 - v_x^2}
	\right\}.	
\end{equation}
The factor $m(v_x,v_y) \, \Delta_x \propto \Delta_x/a$ 
exponentially suppresses large $v_x$ velocities. 
As explained in the main text, this is due to Pauli blocking. 
The corresponding result for the interacting case with $\eta > 0$ is 
\begin{equation}
	\delta n_+(v_x \rightarrow 1,v_y;0,0)  
	\propto
	\frac{\left[m(v_x,v_y) \, v_x\right]^{\eta - 1}}{(1 - v_x^2)^{(3+\eta)/2}}
	\,
	\Phi_x(R_x = 0).	
\end{equation}
In this case, the initial inhomogeneity $\Phi_x(R_x)$ factorizes from the velocity dependence. 
The non-integrable singularity at $v_x = 1$ is due to the Jacobian,
without the exponential suppression that arises in the noninteracting case. 
The singularity is regularized if we retain the ultraviolet scale $\zeta > 0$ in Eq.~(5) 
\cite{S--Foster11}.

\end{document}